\documentstyle[amscd,amssymb,verbatim,12pt]{amsart}
\theoremstyle{plain}

\theoremstyle{definition}

\numberwithin{equation}{section}

\def\dspace{\baselineskip=0.3 in}
\begin{document}
\dspace
\title[Dark Energy and Dark Matter.................]{Dark Energy and Dark Matter of the
  universe from One-loop renormalization of Riccion }

\author[S.K.Srivastava]%
        {    }

\maketitle

\centerline{\bf S.K.Srivastava }

\centerline{ Department of Mathematics, North Eastern Hill University,}

\centerline{ NEHU Campus,Shillong - 793022 ( INDIA ) }

\centerline{e-mail:srivastava@@nehu.ac.in ; sushil@@iucaa.ernet.in }

\vspace{1cm}

\centerline{\bf Abstract}

\smallskip

Here, creation of the universe is obtained only from gravity sector. The dynamical
universe begins with two basic ingredients (i)vacuum energy, also called dark
energy (as vacuum energy is not observed) and (ii) background radiation. These
two are obtained through one-loop renormalization of riccion. Solutions of
renormalization group equations yield initial value of vacuum energy with
density $\rho_{\Lambda_{\rm ew}} =  10^{6} {\rm GeV}^4$ as well as
show a phase transition at the electroweak scale $M_{\rm ew}$. As a result of
phase transition, energy is released (in the form of background radiation) heating the universe upto temperature
$ T_{\rm ew} = 78.5{\rm GeV} = 9.1 \times 10^{14}K $ initially.  In the
proposed cosmology, it is found that not only current universe expands with
acceleration, but it undergoes accelerated expansion from the beginning
itself. It is demonstrated that dark energy decays to dark matter as well as
ratio of dark matter density and dark energy density remains less than unity
upto a long time in future universe also, providing a solution to $ cosmic$
$coincidence$ $ problem.$ Future course of the universe is also discussed
here. It is shown how entropy of the universe grows upto $10^{87}$ in the
present universe. Moreover, particle creation, primordial nucleosynthesis and structure
formation in the late universe is discussed for the proposed model. Thus, investigations, here, present a fresh look to cosmology consistent with current observational evidences as well as provide solution to some important problems.



\vspace{2cm}

\centerline{\bf 1.Introduction}

Experimental probes, like luminosity measurements of high redshift supernova
\cite{ar}, anisotropies in cosmic microwave background \cite{dn} and observational
gravitational clustering \cite{ja}, strongly indicate late time accelerated
expansion of the universe. Theoretically, accelerated expansion of the
universe can be obtained either by modifying left hand side of Einstein's
equations, like introduction of C-field in $steady-$ $state$ $theory$ adhered
to the Perfect Cosmological Principle \cite{fh} or by using energy momentum tensor ,
having dominance of $exotic$ matter, with negative pressure violating the
$strong$ $energy$ $condition$. This kind of matter is known as $dark $
$energy$ (vacuum energy), which has drawn much attention of cosmologists
today. Past few years have witnessed concerted efforts to propose different
$dark $ $energy$ models. Important efforts, in this direction, are scalar
field models like (i) quintessence \cite{br88}, (ii) k-essence \cite{ca},
(iii) tachyon scalar fields \cite{as, mr, ea, gb, ps, az} and models based on
quantum particle production, Chaplygin gas \cite{vs00, vs02}. In these models,
a scalar field acts as a source of are $dark $ $energy$ and plays crucial role
in the dynamical universe. But these models have $no$ answer to  the question
``Where are these scalars coming from?'' It is like Higg's fields in GUTs as
well as inflaton in inflationary models of the early universe.  In the
cosmology probed here, no scalar field is required to be incorporated, from
outside the theory, to discuss cosmic $dark $ $energy.$ Following arguments in
\cite{vs02, brp},here also, vacuum energy is recognized as DE. 

All these dark energy models (mentioned above) try to explain rolling down of
DE density from a very high value in the early universe to an extremely small
value in the current universe as it is suggested by astronomical
observations. But its initial value, in the early universe, is different in
different contexts. For example, it is  $\sim
10^{76} {\rm GeV}^4$ at Planck scale, $\sim 10^{60} {\rm GeV}^4$ at GUT phase
transition and quantum chromodynamics yields its value $\sim 10^{-3} {\rm
  GeV}^4$. In what follows, one-loop renormalization of riccion ( a particle
representing physical role of the Ricci scalar explained below) contributes
initial value equal to  $ 10^{6} {\rm
  GeV}^4$ for slowly varying time-dependent DE density to the dynamical
universe beginning with a phase transition at the electro-weak scale $M_{\rm
  ew} = 100 {\rm GeV}$.

 Work starts effectively from a
general case of $(4 + D)$-dimensional space-time and, later on, it is shown
that $ D \to 6$ making effective dimension of the space-time equal to
$10$. The observable universe is a 4-dimensional hypersurface of the
higher-dimensional world. So far sharpest experiments could probe gravity upto
0.1 mm.and, in the matter sector, probe could be possible  upto electroweak
scale $M_{\rm ew} = 100 {\rm GeV}$, having wavelength as short as $\sim 1.97
\times 10^{-16}$cm. It means that it is difficult to realize 4-dimensional
gravity for length scales less than 0.1 mm. So, it is reasonable to think
higher-dimensional gravity for scales smaller than this scale. This idea is parallel
to the brane-world gravity, where it is assumed that gravity is stronger in
higher-dimensional space-time, called $bulk$ and only a small part of it is
relized in the observable universe \cite{rm}.

It has been noted by many physicists that the $Ricci$ $scalar$ $R$ behaves
like a physical field also, in addition to its geometrical nature, if
gravitational action contains higher-derivative terms \cite{aa, yz, sk93,
  sk94, sk96, sk97, sk98, sk99}. In 1980,
Starobinsky suggested that if sign of the $R^2$ term in the
higher-derivative gravitational action is chosen properly, one could obtain
only one scalar particle with positive energy and positive squared mass. He
called it as ``scalaron''\cite{aa}. In papers \cite{aa, yz},gravitational constant $G$ is
either taken equal to unity or lagrangian density is taken as $\frac{1}{16 \pi
  G} (R +$ higher - derivative terms). As a result, (mass)$^2$ of the Ricci
scalar, does not depend on the gravitational constant $G$. Realizing the
important role of $G$ in gravity, in papers \cite{ sk93,  sk94, sk96, sk97, sk98, sk99} as well as here,
lagrangian density is taken as $(\frac{1}{16 \pi G} R +$ higher - derivative
terms) leading to a drastic change where (mass)$^2$  depends on $G$ also.  In the following
section as well as earlier works , it is shown that physical aspect of the
$Ricci$ $scalar$ is given by a scalar field ${\tilde R} = \eta R$
($\eta$ is a parameter having length dimension), called $riccion$ with
$(mass)^2$ depending on gravitational constant $G$ and other coupling
constants in the action \cite{ sk93,  sk94, sk96, sk97, sk98, sk99}. It is different from $scalaron$ in two ways
(i)mass dimension of $riccion$ is one like other scalar fields ,such as
quintesence, inflaton and Higg's scalar, whereas mass dimension of scalaron is
two and (ii)$(mass)^2$ for scalaron does not depend on $G$, whereas ,for a
riccion, it depends on $G$.

  The present paper begins with higher-derivative gravitational action in higher-dimensional space-time with topology $M^4 \otimes S^D$, where $S^D$ is a $D$-dimensional sphere being the $hidden$ extra-dimensional compact space. Higher-derivative gravity faces ghost problem, which can be avoided by taking coupling constants in the action properly. 
 
The distance function for $(4 + D)$-dimensional space-time  is defined as

$$ d S^2 = g_{\mu\nu} d x^{\mu} d x^{\nu} - l^2 d\Omega^2  \eqno(1.1a)$$
with
$$ d\Omega^2 = d\theta_1^2 + sin^2\theta_1 d\theta^2_2 + \cdots +
sin^2\theta_1 \cdots sin^2\theta_{(D-1)} d\theta^2_D.  \eqno(1.1b)$$
Here  $g_{\mu\nu} (\mu ,\nu = 0, 1, 2, 3)$ are components of the
metric tensor in $M^4$, $l$ is radius of the sphere which is independent of coordinates $x^{\mu}$ and $0 \le \theta_1,\theta_2, \cdots , \theta_{(D-1)} \le \pi$  and $0 \le \theta_D \le 2\pi.$ 

It is important to mention here that $riccion$ is different from the scalar mode of $graviton$. which is highlighted in Appendix A.

Thus, in the proposed cosmological scenario, our dynamical universe begins with the initial $dark$ $energy$ density (vacuum energy density)$\rho_{\Lambda_{\rm ew}} = 1.16 \times 10^{7} {\rm GeV}^4$ and background radiation with temperature $T_{\rm ew} = 33.5 {\rm GeV} = 3.89 \times 10^{14}K$, contributed by $riccion$ at $M_{\rm ew}$. The background radiation is caused by a phase transition at $M_{\rm ew}$. This event is recognized as $big-bang$. As usual, dynamical universe grows in the $post$ $big-bang$ era. In the present theory, $riccion$ has effective role in the $pre$ $big-bang$ era, but it remains $passive$ in the $post$ $big-bang$ times.

The paper is organized as follows. Section 2 demonstrates derivation of
riccion equation and its action. One-loop renormalization of $riccion$ is done
in sestion 3. Renormalization group equations are solved in section 4. This is
an important section as it provides initial values of $dark$ $energy$ density
$\rho_{\Lambda}$ and temperature $T$ of the observable universe. Moreover, in
this section, dimension of $hidden$ space is derived to be $6$. In section 5,
derivation of equation of state for $dark$ $energy$(vacuum energy) is derived
and phase transition at $M_{\rm ew}$ is discussed. It is found that $dark$
$energy$ decreases with  expansion of the universe. So, it is natural to think
for decay of $dark$ $energy$  to  $dark$ $matter$, which is discussed in
section 6. This section is very important from cosmological point of view, as
many important results are derived here. It is demonstrated that $dark$
$energy$ decays to $hot$ $dark$ $matter$ (HDM) till matter remains in thermal
equillibrium with radiation, but when temperature falls down the decoupling
temperature, production of HDM decreases and creation of $cold$ $dark$
$matter$ (CDM)increases. $Dark$ $matter$ density is found less than $dark$
$energy$ density from the epoch of $big-bang$ upto the time $10.6 t_0$ ( $t_0$
is the present age of the universe ). This result provides a solution to $
cosmic$ $coincidence$ $ problem$, which raises the question `` Why does
$dark$ $energy$ dominates over $dark$ $matter$ only recently?.'' First time,
this question was posed by P.J.Steihardt \cite{pj}. In some earlier works also,
this problem was adrressed and solutions were suggested, taking coupled system
of $quintessence$ scalar fields and matter \cite{la, wz, lpc03, lpc04}. As mentioned above,
contrary to earlier attempts, such scalar fields are not required
here. Moreover, it is found that the universe undergoes an accelerated
expansion from the beginning itself upto late future universe. Future course
of dynamics is also discussed here.   In this section, temperature and entropy
of the universe are discussed and it is demonstrated how entropy of the
universe grows upto $10^{87}$ upto the current epoch. In section 7, creation
of spinless and spin-1/2 particles , primordial nucleosynthesis as well as
structure formation in the late universe are discussed. The last section summarizes results.

Natural units, defined as $\kappa_B = \hbar = c = 1$ ( where $\kappa_B$ is Boltzman's constant, $\hbar$ is Planck's constant divided by $2\pi$ and $c$ is the speed of light) are used here with  $\rm GeV$ is used as a fundamental unit such that $1{\rm GeV} = 1.16 \times 10^{13}K = 1.78 \times 10^{-24} gm$ , $ 1{\rm GeV}^{-1} = 1.97 \times 10^{-14}cm = 6.58 \times 10^{-25} sec$.

\bigskip

\centerline{\bf 2.Riccions from (4+D)-dimensional geometry of the space-time}

\smallskip

Theory begins with the gravitational action

$$ S = \int {d^4 x} {d^D y}  \sqrt{- g_{(4+D)}} \quad
\Big[ \frac{M^{(2+D)}R_{(4+D)}}{16 \pi
} + {\alpha_{(4+D)}} R_{(4+D)}^2 + \gamma_{(4+D)} ( R_{(4+D)}^3 $$
$$  - \frac{6(D+3)}{(D-2)}{\Box}_{(D+4)}R^2_{(D+4)})\Big], \eqno(2.1a)$$ 
where  $G_{(4+D)} = M^{- (2+D)}$ ($M$ being the mass scale ),$ \alpha_{(4+D)} = \alpha V_D^{-1},
\gamma_{(4+D)} = \frac{\eta^2}{3! (D-2)} V_D^{-1}, $ and
$ R_D = \frac{D(D-1)}{l^2}.$ Here $V_D$, being the volume of $S^D$ ,is given as
$$ V_D = \frac{2 \pi^{(D+1)/2}}{\Gamma(D+1)/2}l^D. \eqno(2.1b)$$ 
 $g_{(4+D)}$ is  determinant of the metric tensor $g_{MN} (M,N = 0,1,2, \cdots, (3+D)$ and $R_{(4+D)} = R + R_D.$  $\alpha$ is a dimensionless coupling constant, $R$ is the  Ricci scalar in $M^4$ and $G_N = G_{(4+D)}/V_D.$ 

Invariance of $S$ under transformations $g_{MN} \to  g_{MN}
+ {\delta} g_{MN}$ yields \cite{sk98,sk99}

$$ \frac{M^{(2+D)}}{16 \pi} (R_{MN} - {1 \over 2} g_{MN} R_{(4+D)} ) + {\alpha_{(4+D)}} H^{(1)}_{MN} + {\gamma_{(4+D)}} H^{(2)}_{MN} = 0,\eqno(2.2a)$$ 
where

$$ H^{(1)}_{MN} = 2 R_{; MN} - 2 g_{MN} {\Box}_{(4+D)} R_{(4+D)} - {1\over 2} g_{MN} R^2_{(4+D)} + 2 R_{(4+D)} R_{MN}, \eqno(2.2b)$$  
and
$$ H^{(2)}_{MN} = 3 R^2_{; MN} - 3 g_{MN} {\Box}_{(4+D)} R^2_{(4+D)} - \frac{6(D+3)}{(D-2)}\{- { 1 
\over 2} g_{MN}{\Box}_{(4+D)} R^2_{(4+D)}$$  $$+ 2{\Box}_{(4+D)}R_{(D+4)}R_{MN} + R^2_{; MN}\} - { 1 \over 2} g_{MN} R^3_{(4+D)} + 3 R^2_{(4+D)} R_{MN}  \eqno(2.2c) $$
with semi-colon (;) denoting curved space covariant derivative and
$${\Box}_{(4+D)} = {1\over \sqrt{-g_{(4+D)}}}{\frac{\partial}{\partial x^M}}\Big( \sqrt{-g_{(4+D)}}\quad g^{MN} \frac{\partial}{\partial x^N}\Big)$$

Trace of these field equations is obtained as

$$- \Big[\frac{(D+2)M^{(2+D)}}{32 \pi }\Big] R_{(4+D)}  +   {\alpha_{(4+D)}} [2(D+3) {\Box}_{(4+D)} R_{(4+D)}  +  {1 \over2} D R^2_{(4+D)} ] $$
$$ + {1 \over2}\gamma_{(4+D)} (D - 2)R^3_{(4+D)}  = 0 .   \eqno(2.3)$$

In the space-time described by the distance function defined in eq.(1.1),

$$ {\Box}_{(4+D)} R_{(4+D)} = {\Box} R = {1\over \sqrt{-g}}{\frac{\partial}{\partial x^{\mu}}}\Big( \sqrt{-g}\quad g^{\mu\nu} \frac{\partial}{\partial x^{\nu}}\Big) R ,   \eqno(2.4)$$
using the definition of $R_{(4+D)}$ given in eq.(2.1).

Connecting eqs.(2.3)- (2.4) as well as using $R_{(4+D)}, \alpha_{(4+D)}$  and $\gamma_{(4+D)}$ from eq.(2.1), one obtains in $M^4$
$$- \Big[\frac{(D+2)M^{(2+D)}V_D}{32 \pi }\Big](R + R_D)  +   \alpha [2(D+3)
{\Box} R $$
$$ +  {1 \over2} D (R + R_D)^2 ]  + \frac{\eta^2}{12}(R + R_D)^3  = 0 ,
\eqno(2.5)$$ 
 which is re-written as

 $$[{\Box} + {1 \over2} \xi R + m^2 + \frac{\lambda}{3!} \eta^2 R^2] R  + \eta^{-1} \vartheta = 0 ,  \eqno(2.6)$$
 where 

 \begin{eqnarray*}
 \xi & = & \frac{D}{2(D+3)} +  \eta^2 \lambda R_D\\ m^2  & = & - \frac{(D + 2) \lambda M^{(2+D)}V_D}{16 \pi} + \frac{D R_D}{2(D+3)} + \frac{1}{2} \eta^2 \lambda R_D^2 \\ \lambda  & = & \frac{1}{4(D+3)\alpha}, \\ \vartheta & = & \eta \Big[- \frac{(D + 
2) \lambda M^{(2+D)}V_D}{16 \pi} + \frac{D  R_D^2}{4 (D+3)} + \frac{1}{6} \eta^2 \lambda R_D^3 \Big],
 \end{eqnarray*} 

 $$   \eqno(2.7 a,b,c,d)$$
 where $\alpha > 0$ to avoid the ghost problem.

A scalar field, representing a spinless particle, has unit mass dimension in
existing theories. $R$, being combination of second order derivative as well
as squares of first order derivative of metric tensor components with respect
to space-time coordinates, has mass dimension 2. So, to have  mass dimension
like other scalar fields, eq.(2.6) is multiplied by $\eta$ and $\eta R$ is
recognized as $\tilde R$.
 As a result, this equation looks like

 $$[{\Box} + {1 \over2} \xi R + m^2 + \frac{\lambda}{3!} {\tilde R}^2]{\tilde R} + \vartheta  = 0.  \eqno(2.8)$$
 For $scalaron$, $\eta$ is dimensionless \cite{aa}.

The above analyses show that, on taking trace of field equations (2.2) and compactifying the space-time $M^4 \otimes S^D$ to $M^4$, $only$ $one$ $degree$ of $freedom$ is obtained spontaneously, which is the scalar mode $\tilde R$ [16, 18-23]. It is unlike $gravitons$ having 5 degrees of freedom including one scalar. In Appendix A, it is explained that scalar mode of $graviton$ is $different$ from $riccion$.

If ${\tilde R}$ is a basic physical field, there should be an action $S_{\tilde R}$ yielding eq.(2.8) for invariance of $S_{\tilde R}$, under transformations ${\tilde R} \to {\tilde R} + \delta {\tilde R}$.

In what follows, $S_{\tilde R}$ is obtained. If such an action exists, one can write
$$\delta S_{\tilde R} = - \int {d^4 x} \sqrt{- g} \delta{\tilde R}[({\Box} + {1 \over2} \xi R + m^2 + \frac{\lambda}{3!} {\tilde R}^2) {\tilde R} + \vartheta] \eqno(2.9a)$$
which yields eq.(2.8) if $\delta S_{\tilde R} = 0$ under transformations ${\tilde R} \to {\tilde R} + \delta {\tilde R}.$

Eq.(2.9a) is re-written as
\begin{eqnarray*}
\delta S_{\tilde R} & = &  \int {d^4 x} \sqrt{- g} \Big[ 
\partial^{\mu}{\tilde R} \partial_{\mu}(\delta{\tilde R}) -  \Big({1 \over2 } \xi R {\tilde R}^2 + m^2 {\tilde R}+ \frac{\lambda}{3!} {\tilde R}^3 + \vartheta
\Big)\delta{\tilde R} \Big]\\ & = &  \int {d^4 x} \delta \Big\{\sqrt{- g}\Big[{1 \over2}  \partial^{\mu}{\tilde R}
\partial_{\mu}{\tilde R} 
-  \Big({1 \over 3! } \xi R{\tilde R}^2 + {1 \over 2} m^2 {\tilde R}^2 +
\frac{\lambda}{4!} {\tilde R}^4 + \vartheta {\tilde R}\Big)\Big]\Big\}.
\end{eqnarray*}
$$ \eqno(2.9b)$$

$R, {\tilde R}$ and ${d^4 x} \sqrt{- g}$ are invariant under co-ordinate transformations. So, $R(x) = R(X), {\tilde R}(x) = {\tilde R}(X)$ and ${d^4 x} \sqrt{- g} = d^4 X,$ where $X^i (i=0,1,2,3)$ are local and  $x^i (i=0,1,2,3)$ are global coordinates. Moreover,

$${\Box} = g^{ij} \frac{\partial^2}{\partial x^i \partial x^j} + \frac{1}{2}  g^{mn} \frac{\partial g_{mn}}{\partial x^i} g^{ij}\frac{\partial}{\partial x^j} + \frac{\partial g^{ij}}{\partial x^i}\frac{\partial }{\partial x^j} = \frac{\partial^2}{\partial X^i \partial X^j} $$
in a locally inertial co-ordinate system, where $g_{ij} = \eta_{ij} $ (components of Minkowskian metric) and $g^{ij}_{,i} = 0$ (comma (,) stands for partial derivative). Thus, in a locally inertial co-ordinate system,
\begin{eqnarray*}
\delta S_{\tilde R} & = & \int {d^4 x} \delta \Big\{\sqrt{- g}\Big[{1 \over2}  \partial^{\mu}{\tilde R}\partial_{\mu}{\tilde R} -  \Big({1 \over 3! } \xi R {\tilde R}^2 + {1 \over 2} m^2 {\tilde R}^2 + \frac{\lambda}{4!} {\tilde R}^4 + \vartheta {\tilde R}\Big) \Big]\Big\} \\ & = & \int {d^4X}\quad \delta \Big\{\sqrt{- g}\Big[{1 \over2}  \partial^{\mu}{\tilde R}
\partial_{\mu}{\tilde R} 
-  \Big({1 \over 3! } \xi R {\tilde R}^2 + {1 \over 2} m^2 {\tilde R}^2 + \frac{\lambda}{4!} {\tilde R}^4 + \vartheta {\tilde R}\Big) \Big]\Big\} \\ & = & \delta \int {d^4X}\quad  \Big\{\sqrt{- g}
\Big[{1 \over2}  \partial^{\mu}{\tilde R}
\partial_{\mu}{\tilde R} -  \Big({1 \over 3! } \xi R {\tilde R}^2  + {1 \over 2} m^2 {\tilde R}^2 + 
\frac{\lambda}{4!} {\tilde R}^4 + \vartheta {\tilde R}\Big)\Big]\Big\}  
\end{eqnarray*}

Employing principles of covariance and equivalence as well as eq.(2.9b)
$$\delta S_{\tilde R} = \delta \int {d^4 x} \Big\{\sqrt{- g}\Big[{1 \over2}  \partial^{\mu}{\tilde R}
\partial_{\mu}{\tilde R} 
-  \Big({1 \over 3! } \xi R {\tilde R}^2 + {1 \over 2} m^2 {\tilde R}^2 + 
\frac{\lambda}{4!} {\tilde R}^4 + \vartheta {\tilde R}\Big) \Big]\Big\}, $$ 
which implies that
 $$ S_{\tilde R} =\int {d^4 x}  \Big\{\sqrt{- g}
\Big[{1 \over2}  \partial^{\mu}{\tilde R}
\partial_{\mu}{\tilde R} 
-  \Big({1 \over 3! } \xi R {\tilde R}^2 + {1 \over 2} m^2 {\tilde R}^2 + 
\frac{\lambda}{4!} {\tilde R}^4 + \vartheta {\tilde R}\Big) \Big]\Big\}. \eqno(2.10) $$

It is important to mention here that  ${\tilde R}$ is different from other scalar fields due to dependence of (mass)$^2$ on the gravitational constant, dimensionality of the space-time and the 
coupling constant $\alpha$, given by the eq.(2.7b). Moreover, it emerges from  geometry of the space-time.

\bigskip

\centerline{\bf 3.One-loop quantum correction and renormalization  of riccion}

\smallskip

The $S_{\tilde R} $ with the lagrangian density, given by eq.(2.10), can be expanded around the classical minimum ${\tilde R}_0$ in powers of quantum fluctuation ${\tilde R}_q = {\tilde R} - {\tilde R}_0 $ as

$$ S_{\tilde R} =  S_{\tilde R}^{(0)} + S_{\tilde R}^{(1)} +  S_{\tilde R}^{(2)} + \cdots ,$$
where
\begin{eqnarray*}
S_{\tilde R}^{(0)} & = &  \int {d^4 x}  \sqrt{- g} 
\Big[{1 \over2} g^{\mu\nu} \partial_{\mu}{\tilde R}_0 
\partial_{\nu}{\tilde R}_0 -  \Big({1 \over 3! \eta} \xi {\tilde R}^3_0 + {1 \over 2} m^2 {\tilde R}^2_0 + 
\frac{\lambda}{4!} {\tilde R}^4_0 + \vartheta {\tilde R}_0 \Big) \Big] \\ S_{\tilde R}^{(2)} & = &  \int 
{d^4 x}  \sqrt{- g} {\tilde R}_q[{\Box} + \frac{1}{2} \xi 
R + m^2 + \frac{\lambda}{2!} {\tilde R}^2_0] {\tilde R}_q
\end{eqnarray*}
and

$$ S_{\tilde R}^{(1)} = 0$$ 
as usual, because this term contains the classical equation.

The effective action is expanded in powers of $\hbar$ (with
$\hbar = 1$) as 
$$ \Gamma({\tilde R}) = S_{\tilde R} + \Gamma^{(1)} + \Gamma^{\prime}$$
with one-loop correction given as \cite{ndb, skp}

$$ \Gamma^{(1)} = {i \over 2} ln Det (D/{\mu}^2),   \eqno(3.1a)$$
where

$$ D \equiv \frac{{\delta}^2 S_{\tilde R}}{{\delta}{\tilde R}^2}
\Big|_{{\tilde R}={\tilde R}_0} = {\Box} + \frac{1}{2} \xi
R + m^2 + \frac{\lambda}{2!} {\tilde R}^2_0   \eqno(3.1b)$$
and  $\Gamma^{\prime}$  is  a term for higher-loop quantum corrections. In eq.(3.1), $\mu$ is a mass parameter to keep $\Gamma^{(1)}$ dimensionless.

To evaluate $\Gamma^{(1)}$,  the  operator regularization method \cite{rb} is used upto adiabatic order 4. As potentially divergent terms are expected upto this order only. In a 4-dim. theory, one-loop correction is obtained as

\begin{eqnarray*}
\Gamma^{(1)} & = & (16 {\pi}^2 )^{-1} \frac{d}{d s} \Big[ \int {d^4 x}\sqrt{-g (x)} 
\Big(\frac{{\tilde M}^2}{{\mu}^2}\Big)^{-s} \Big\{\frac{{\tilde M}^4}{(s - 2)(s - 1)} \\ && + \frac{{\tilde M}^2}{(s - 1)}\Big(\frac{1}{6} - \frac{1}{2} \xi 
\Big) R + \Big[ {1 \over 6} \Big(\frac{1}{5} - \frac{1}{2} \xi \Big){\Box} R + {1 \over 180} 
R^{\mu\nu\alpha\beta} R_{\mu\nu\alpha\beta}\\ && 
-{1 \over 180} R^{\mu\nu} R_{\mu\nu} + {1 \over 2}\Big(\frac{1}{6} - \frac{1}{2} \xi \Big)^2 R^2 
\Big]\Big\}\Big] \Big|_{s=0},
\end{eqnarray*}
$$   \eqno(3.2a)$$
where $$ {\tilde M}^2 = m^2  +  (\lambda/2) {\tilde R}^2_0 .  \eqno(3.2b)$$

Here it is important to note that matter as well as geometrical both aspects of the Ricci scalar are used in eq.(3.2). The matter aspect is manifested by ${\tilde R}$ and the geometrical aspect by $R,$ Ricci tensor components  $R_{\mu\nu}$ and curvature tensor components $R_{\mu\nu\alpha\beta}$ as it is mentioned above also.

After some manipulations, the lagrangian density in ${\Gamma}^{(1)}$ is obtained as 

\begin{eqnarray*}
L_{\Gamma^{(1)}} & = & (16 {\pi}^2 )^{-1}  \Big[ (m^2  +  (\lambda/2) {\tilde R}^2_0 )^2 \Big\{{3 \over 4} - {1 \over 2} ln \Big(\frac{m^2  +  (\lambda/2) {\tilde R}^2_0 
}{{\mu}^2} \Big)\Big\} \\ &&  - \Big({1 \over 6} - \frac{1}{2} \xi \big)  R ( m^2  + 
(\lambda/2) {\tilde R}^2_0 ) \Big\{1 - ln \Big(\frac{m^2  +  (\lambda/2) {\tilde 
R}^2_0}{{\mu}^2} \Big)\Big\} \\ &&- ln \Big(\frac{m^2  +  (\lambda/2) {\tilde R}^2_0}{{\mu}^2} \Big)\Big\{ {1 \over 6}\Big(\frac{1}{5} - \frac{1}{2} \xi \Big) {\Box}R + \frac{1}{180} R^{\mu\nu\alpha\beta} R_{\mu\nu\alpha\beta} \\&& -{1 \over 180} R^{\mu\nu} R_{\mu\nu} + {1 \over 2}\Big(\frac{1}{6} - \frac{1}{2} \xi \Big)^2 R^2 \Big\}\Big] .
\end{eqnarray*}

$$   \eqno(3.3)$$

Now the renormalized form of lagrangian density can be written as

\begin{eqnarray*}
L_{\rm ren} &=& {1 \over 2} g^{\mu\nu} {\partial}_{\mu} {\tilde R}_0 {\partial}_{\nu} {\tilde R}_0  - \frac{\xi}{3! \eta} {\tilde R}_0^3- {1 \over 2}m^2 {\tilde R}^2_0  - {\lambda \over 4!} {\tilde R}^4_0 - \vartheta {\tilde R}_0 + {\tilde \Lambda} \\ &&  + {\epsilon}_0 R + {1 \over 2} 
{\epsilon}_1 R^2 + {\epsilon}_2 R^{\mu\nu} R_{\mu\nu} + {\epsilon}_3 R^{\mu\nu\alpha\beta} R_{\mu\nu\alpha\beta} \\&& + {\epsilon}_4 {\Box} R + L_{\Gamma^{(1)}} + L_{\rm ct}
\end{eqnarray*}

$$  \eqno(3.4a)$$
with bare coupling constants $\lambda_i \equiv ( m^2, \vartheta,\lambda, {\tilde \Lambda}, \xi, \epsilon_0, \epsilon_1, \epsilon_2, \epsilon_3, \epsilon_4 ), \Gamma^{(1)} $ 
given by eq.(3.3) and $L_{\rm ct}$ given as

\begin{eqnarray*}
L_{\rm ct} &=&  - {1 \over 2} \delta\xi R {\tilde R}^2_0 - {1 \over 2}\delta m^2 {\tilde R}^2_0  - 
{\delta \lambda \over 4!} {\tilde R}^4_0 - \delta \vartheta {\tilde R}_0 + \delta{\tilde \Lambda}  + 
{\delta\epsilon}_0 R + {1 \over 2}{\delta\epsilon}_1 R^2 \\ && + {\delta\epsilon}_2 R^{\mu\nu} R_{\mu\nu} + {\delta\epsilon}_3 
R^{\mu\nu\alpha\beta} R_{\mu\nu\alpha\beta} + {\delta\epsilon}_4 {\Box} R .
\end{eqnarray*}
$$  \eqno(3.4b)$$

In eq.(3.4b), $\delta\lambda_i \equiv (\delta m^2, \delta\vartheta, \delta
\lambda, \delta{\tilde \Lambda}, \delta\xi, \delta\epsilon_0,
\delta\epsilon_1, \delta\epsilon_2, \delta\epsilon_3, \delta \epsilon_4 )$ are
counter-terms, which are calculated using the following renormalization
conditions \cite{blh, ee}

\begin{eqnarray*}
{\tilde \Lambda} & = &  L_{\rm ren} |_{{\tilde R}_0 = {\tilde R}_{(0)0} , R=0}\\ 
\lambda & = &  - \frac{{\partial}^4}{\partial {\tilde R}^4_0} L_{\rm ren}
\Big|_{{\tilde R}_0} = {\tilde R}_{(0)1 , R=0} \\ \vartheta  & = & - \frac{\partial}{\partial {\tilde R}_0} L_{\rm ren}
\Big|_{{\tilde R}_0 = {\tilde R}_{(0)1} , R=0} \\
 m^2 & = &  - \frac{{\partial}^2}{\partial {\tilde R}^2_0} L_{\rm ren}
\Big|_{{\tilde R}_0 = 0 , R=0} \\
\frac{1}{2} \xi & = &  - \eta \frac{{\partial}^3}{{\partial R}{\partial {\tilde R}^2_0}} L_{\rm ren} 
\Big|_{{\tilde R}_0 = {\tilde R}_{(0)2} , R=0} \\
\epsilon_0 & = &  \frac{\partial}{\partial R} L_{\rm ren}
\Big|_{{\tilde R}_0 = 0 , R=0} \\
\epsilon_1 & = &  \frac{{\partial}^2}{\partial R^2} L_{\rm ren}
\Big|_{{\tilde R}_0 = 0 , R = R_5} \\
\epsilon_2 & = &   \frac{\partial}{\partial( R^{\mu\nu}R_{\mu\nu})} L_{\rm ren}
\Big|_{{\tilde R}_0 = 0 , R = R_6} \\
\epsilon_3 & = &   \frac{\partial}{\partial(R^{\mu\nu\alpha\beta
}R_{\mu\nu\alpha\beta})} L_{\rm ren}
\Big|_{{\tilde R}_0 = 0 , R = R_7} \\
\epsilon_4 & = &  \frac{\partial}{\partial ({\Box} R)} L_{\rm
ren}\Big|_{{\tilde R}_0 = 0 , R = R_8} .
\end{eqnarray*}
$$  \eqno(3.5a,b,c,d,e,f,g,h,i,j)$$
\noindent As ${\tilde R} = {\eta} R,$ so when $R = 0, {\tilde R}_{(0)0} = {\tilde R}_{(0)1} = 
{\tilde R}_{(0)2} = 0 \quad {\rm  and}\quad  R_5 = R_6 = R_7 = R_8 = 0 \quad {\rm when}\quad {\tilde R}_0 = 0 .$

\bigskip

\centerline {\bf 4.Renormalization group equations and their solutions}

\smallskip

Mass scale dependence of coupling constants are obtained by solving renormaliztion group equations

$$ \frac{d {\lambda}_i}{d\tau} = \beta_{\lambda_i},  \eqno(4.1)$$
where $ \tau = {1 \over 2} ln (M^2_{\rm ew}/{\mu}^2)$  and $\beta_{\lambda_i}$ are one-loop $\beta$ functions for different coupling constants. Here $\mu$ is the mass scale parameter  and $M_{\rm ew}$ is the cut-off scale such that $\mu \ge M_{\rm ew}.$ As experimental probes could be possible upto  $M_{\rm ew}$, so this scale is used as a cut-off mass scale.

$\beta_{\lambda_i}$ in eq.(4.1) are calculated using counter-terms yielded by eqs.(3.4) and (3.5) and putting $\mu \frac{d}{d\mu}\lambda_{i} = 0$ for bare coupling constants in equations

$$ \beta_{\lambda_i} = \mu \frac{d}{d\mu}(\lambda_{i} + 
\delta\lambda_i)\Big|_{\lambda_i}   \eqno(4.2)$$

Using $\beta$-functions for different coupling constants, given by
eqs.(4.2), solutions of differential equations (4.1) are derived,  as

\begin{eqnarray*}
\tilde\Lambda & = & \tilde\Lambda_{\rm ew} + \frac{m^4_{\rm ew}}{2\lambda_{\rm ew}} \Big[\Big(1 - \frac{3
\lambda_{\rm ew}\tau}{8 \pi^2}\Big)^{1/3} - 1 \Big] \\
\lambda & = & \lambda_{\rm ew} \Big[1 - \frac{3 \lambda_{\rm ew} \tau}{8 \pi^2} \Big]^{-1}\\ \vartheta & = & \vartheta_{\rm ew} (constant) \\
 m^2 & = & m^2_{\rm ew} \Big[1 - \frac{3 \lambda_{\rm ew}\tau}{8 \pi^2} \Big]^{-1/3}\\
\frac{1}{2} \xi & = &  {1 \over 6}  + \Big(\frac{1}{2} \xi_{\rm ew} - \frac{1}{6} \Big) \Big[1 - \frac{3
\lambda_{\rm ew} \tau}{8 \pi^2} \Big]^{-1}
\end{eqnarray*}
$$  \eqno(4.3a,b,c,d,e)$$
for coupling constants of relevant terms to be used in further investigations in the present paper. Here, $\lambda_{i_{\rm ew}} = \lambda_i (\tau=0)\quad {\rm and}\quad  \tau = 0 \quad {\rm at}\quad  \mu = M_{\rm ew} $ according to defnition of $\tau$ given above.

These results show that as $\mu \to \infty ( \tau \to  - \infty ),\quad \lambda \to  0 \quad {\rm and}\quad m^2 \to 0$ and $\frac{1}{2} \xi \to \frac{1}{6}.$ 

Using these limits in eq.(2.7a), it is obtained that

 $$ D =  6.  \eqno(4.4 )$$

Also,eqs.(2.1a) and (4.4) imply

 $$ R_6 = \frac{30}{l^2} .   \eqno(4.5)$$

So, from eq.(4.3a)
$$\tilde\Lambda =  \tilde\Lambda_{\rm ew}   \eqno(4.6)$$
at $\mu = M_{\rm ew} $. The equation (4.3c) shows that $\vartheta$ is
independent of mass scale $\mu$. So, $\eta^{-1} \vartheta$, being true for all
$\mu$, is obtained from eq.(2.7d) as
$$\eta^{-1}\vartheta  = \frac{30}{l^2_{\rm ew}}\Big[m^2_{\rm ew} - \frac{5 ( 1 + 60 \lambda_{\rm ew})}{l^2_{\rm ew}} \Big], \eqno(4.6)$$
at $\mu = M_{\rm ew}$. Here
$$  m^2_{\rm ew} = - \frac{\lambda_{\rm ew} M^8_{\rm ew} V_{6{\rm ew}}}{2 \pi} + \frac{10 ( 1 + 450 \lambda_{\rm ew})}{l^2_{\rm ew}} \eqno(4.7)$$
with 
$$ V_{6{\rm ew}} = \frac{16 \pi^3 l^6_{\rm ew}}{15} \eqno(4.8)$$
which is derived connecting eqs.(2.7b,d), taking the arbitrary parameter $\eta
= l_{\rm ew}$ and putting $D = 6.$  

Thus $\eta^{-1}\vartheta$ , given by eq.(4.6),is an imprint of extra
six-dimensional compact manifold $S^6$ in the 4-dimensional universe. This
term has the dimension of energy density. Moreover, it is not generated
through matter, but the geometry of extra-dimensional space. So, it is
recognized as $dark$ $energy$ density $\rho_{\Lambda_{\rm ew}}$ at $\mu = M_{\rm
  ew}$, given as
$$\rho_{\Lambda_{\rm ew}} = \eta^{-1}\vartheta  = \frac{30}{l^2_{\rm ew}}\Big[m^2_{\rm ew} - \frac{5 ( 1 + 60 \lambda_{\rm ew})}{l^2_{\rm ew}} \Big], \eqno(4.9)$$

\centerline {\bf 5.  Equation of state for dark energy and Phase transition  }

\centerline{\bf   at the eletroweak scale }

$Dark$ $energy$ density $\rho_{\Lambda_{\rm ew}}$,
obtained through renormalization of riccion at energy mass scale $\mu = M_{\rm
  ew}$ , is given by eq.(4.9). As riccion is emerging from geometry of the
space-time,  $\rho_{\Lambda_{\rm ew}}$ can be obtained as zero-point energy of riccion also, given as

$$ \rho_{\Lambda} (t=0) = \rho_{\Lambda_{\rm ew}} = (2 \pi)^{-3} \int_0^k \sqrt{k^2 + m^2_{\rm ew}}4 \pi k^2 {dk},  \eqno(5.1)$$
where $m_{\rm ew}$ is the mass of riccion at the electroweak scale $M_{\rm   ew}$. Eq.(5.1) shows that $\rho_{\Lambda_{\rm ew}}$ diverges  as $k \to \infty.$ But $\rho_{\Lambda_{\rm ew}} =  = \tilde\Lambda_{\rm ew} $ (as given by eq.(4.7)), being
finite, implies that the integral in eq.(5.1) should be regularized upto  a
certain cut-off mode $k = k_c$. As so far experiments could be performed upto
$M_{\rm ew}$ only, so cut-off scale is taken as $k_c = M_{\rm ew}$ as
above. Moreover, at this scale riccions heavier than $M_{\rm ew}$ can not
survive, so $m_{\rm ew} \le M_{\rm ew}$. Now eq.(5.1) yields

$$\rho_{\Lambda_{\rm ew}} = \frac{\pi}{4(2 \pi)^3}M^4_{\rm ew}[3 \sqrt{2} -
ln(1 + \sqrt{2})]   \eqno(5.2)$$
taking $m_{\rm ew} = M_{\rm ew}$.
The second quantization and uncertainty relation imply that vacuum has energy
density as well as pressure \cite{vs00}. Experimental probes, like Ia supernova and 
WMAP \cite{ar, dn, ja, abl} suggest accelerated expansion of the universe, which requires negative pressure for the vacuum. So, to have consistency with recent probes, the isotropic vacuum pressure is calculated as

$$p_{\Lambda_{\rm ew}} = - \frac{(2 \pi)^{-3}}{3} \int_0^{m_{\rm ew}}4 \pi k^3 {dk} = - \frac{ \pi}{3(2 \pi)^3} M^4_{\rm ew} . \eqno(5.3)$$
 It yields 

$$p_{\Lambda_{\rm ew}}/\rho_{\Lambda_{\rm ew}} = \omega_{\Lambda_{\rm ew}} =
- \frac{4}{3}[3 \sqrt{2} -
ln(1 + \sqrt{2})]^{-1} = - 0.397 \simeq - 0.4.   \eqno(5.4a)$$

Though in certain models, time-dependence of $\omega = p/\rho$ is also proposed , but normally it is taken as a constant. So, here also, this ratio is considered independent of time. As a result

$$\omega_{\Lambda} = p_{\Lambda}/\rho_{\Lambda} = p_{\Lambda_{\rm ew}}/\rho_{\Lambda_{\rm ew}} = - 0.4 .  \eqno(5.4b)$$

Using $l^{-1}_{\rm ew} = M_{\rm ew} = m{\rm ew} $ and connecting
eqs.(4.7),(4.8) and (5.2), a quardatic equation for $\lambda{\rm ew}$ is
obtained as 

$$\Big(450 - \frac{8 \pi^2}{15} \Big)\lambda{\rm ew}^2 + \Big[ 20\Big(450 -
\frac{8 \pi^2}{15} \Big) + 30\times 2818.8 \Big]\lambda{\rm ew} + 11375.3 = 0, \eqno(5.5)$$   
 which is solved to
$$\lambda_{\rm ew} = -0.013352958 .   \eqno(5.6)$$

Also using $M_{\rm ew} = m{\rm ew} $ in eq.(5.2), it is obtained that

$$\rho_{\Lambda_{\rm ew}} \simeq 10^{6} {\rm GeV}^4.  \eqno(5.7)$$

Planck scale is supposed to be a fundamental scale in field theories. So, it
is proposed that energy mass scale $\mu$ falls from the Planck mass $M_P = 10^{19}{\rm GeV} $ to the cut-off scale $M_{\rm ew} = 100  {\rm GeV} $. When it
  happens so, phase transition takes place at $\mu = M_{\rm ew}$ and energy
  with density
$$ \rho_{{\rm ew}(r)} =  \tilde \Lambda - \tilde \Lambda_{\rm ew} = 2.5 \times 10^7 {\rm GeV}^4   \eqno(5.8a)$$ 
is released, which is obtained connecting eqs.(4.3a) and (5.6).

 In the proposed model (PM), this event is recognized as $big-bang$, being beginning of the universe like standard model(SMU) with the release of
$background$ $radiation$ having energy density $\rho_{{\rm ew}(r)}$, given by
eq.(5.8a). In contrast to SMU, here, released energy density at the epoch of
big-bang is finite (it is infinite in SMU).  Temperature of photons $T_{\rm ew}$ with energy density $\rho_{\rm ew}$, is obtained from

$$ \rho_{{\rm ew}(r)} =  \frac{  \pi^2}{15} T^4_{\rm ew} = 2.5 \times 10^7 {\rm GeV}^4   \eqno(5.8b)$$
as

$$ T_{\rm ew} = 78.5{\rm GeV} = 9.1 \times 10^{14}K .  \eqno(5.9)$$

\bigskip

\centerline {\bf 6.Proposed cosmological scenario, Dark energy and Dark matter }

\smallskip

In the standard model of the $big-bang$ theory, it is supposed that, around
$13.7 {\rm Gyrs}$ ago, there used to be a $fireball$ ( an extremely hot
object), which was termed as $primeval$ $atom$ by Lema$\hat i$tre. Our
universe came into existence, when this $primeval$ $atom$ burst out. This event is called $big-bang$.

According to the proposed cosmological picture, the main content of the
$primeval$ $atom$ was riccions,being
contribution of 10-dimensional higher-derivative gravity to the 4-dimensional
world. One-loop renormalization of $riccions$ and solutions of
resulting group equations yield that when energy mass scale comes down to
electroweak scale $M_{\rm ew}= 100 {\rm GeV}$, $riccion$ contributes $dark$
$energy$  density $\rho_{\Lambda_{\rm ew}} = 10^{6} {\rm GeV}^4$. Moreover,  phase transition takes place at
$M_{\rm ew}$, releasing the background radiation. This radiation thermalizes
the universe upto the temperature $T_{\rm ew} = 9.1 \times 10^{14}K$. Here,
the event of phase transition is recognized as $big-bang$, which heralds our
dynamical universe having the initial temperature $T_{\rm ew} = 9.1 \times
10^{14}K$ and initial value of dark energy density $\rho_{\Lambda_{\rm ew}} =
10^{6} {\rm GeV}^4$. 

Moreover, it is important to mention that existence of $riccions$ are possible
at energy scales where higher-derivative terms in the gravitational action
(2.1a) has significant role compared to Einstein-Hilbert term. At energy
scales below $M_{\rm ew}$, Einstein-Hilbert term dominates higher-derivative
terms in the action (2.1a). So, $riccions$ has no direct role in the evolution
of proposed model of the universe, but it has two very important contributions
to the observable universe as  value of dark energy density
$\rho_{\Lambda_{\rm ew}} = 10^{6} {\rm GeV}^4$ and cosmic background radiation
temperature $T_{\rm ew} = 9.1 \times 10^{14}K$ at cosmic time $t=0.$

Astronomical observations have compelling evidences that the current universe
is dominated by dark energy(DE). The present cosmic dark energy density is
very low, but it used to be very high in the early universe. The fall of dark
energy density from very high to extremely low value can be explained if it is
time-dependent. So, like other other cosmic dark energy models, here also,
dark energy density $\rho_{\Lambda}$ is slowly varying function of
time. Following Bronstein's idea \cite{vs02, brp, bro} that DE could decay to
hot or cold dak matter, here it is proposed that $\rho_{\Lambda}(t)$ decays to
hot dak matter(HDM) till temperature is high and cold dark matter after
decoupling of matter from radiation. As a result, $\rho_{\Lambda}(t)$ falls
from high value $10^{6} {\rm GeV}^4$ to currently low value $0.73 \rho_{{\rm
    cr},0}$ (where $\rho_{{\rm cr},0}$ is the critical density).

 It is proposed that the homogeneous dynamical universe begins with topology having the distance function 

$$ dS^2 = dt^2 - a^2(t) [ dx^2 + dy^2 + dz^2 ] \eqno(6.1)$$ 
for the spatially flat model of the universe supported by recent
experiments [1, 2, 3, 34]. This space-time is a special case of hypersurface $M^4$ of the line element (1.1). Here $a(t)$ is the scale factor.

In what follows, it is obtained that $dark$ $ energy$ $ density$ decreases
with time from its initial value $ 10^{6} {\rm GeV}^4$ and falls down by 53
orders in the present universe. In 1933, Bronstein proposed that, in the
expanding universe, $dark$ $ energy$ $ density$ decreases due its decay as a
result of emission of dark matter or radiation \cite{brp, bro}. The radiation,
so emitted, could disturb spectrum of 3K-microwave background radiation. So,
Bronstein's original idea was modified and it was introduced that dark energy
(DE) could decay to hot or cold dark matter without any harm to spectrum of
3K-microwave background radiation \cite{brp}. Following this idea, here,   it is demonstrated that $dark$ $ energy$ decays to $dark$ $ matter$ providing a solution to $cosmic$ $ coincidence$ $ problem$.

It is shown above that , in the beginning, the universe was very hot due to the background radiation (released during phase transition). Radiation energy density falls as

$$\rho_r = \frac{\rho_{{\rm ew}(r)} a^4_{\rm ew}}{a^4(t)}. \eqno(6.2)$$
with growing scale factor.

Matter remains in thermal equillibrium with the background radiation for sufficiently long time. According to WMAP \cite{abl}, decoupling of matter from background radiation takes place at $t_d \simeq 386 kyr = 1.85 \times 10^{37}{\rm GeV}^{-1}$. Equation of state for radiation is $\omega = 1/3$. So, here, it is proposed that $dark$ $ energy$ decays to HDM till decoupling time $t_d$ obeying $\omega_m = 1/3$.

When $ t > t_d$, it decays more to CDM ( which is non-baryonic and pressureless) with $\omega_m = 0$ as well as HDM. So, ratio of densities of HDM and CDM is extremely small below $t_d$.

In what follows, development of the universe is probed using these ideas.

\bigskip

\noindent {\bf (a) \underline{Decay of $dark$ $ energy$ to $dark$ $ matter$}}

\smallskip

The conservation equations

$$ T^i_{(\Lambda) j;i} + T^i_{({\rm dm}) j;i} = 0    \eqno(6.3)$$ yield
coupled equations

$${\dot \rho}_{\Lambda} + 3 H ( 1 + {\omega}_{\Lambda}) \rho_{\Lambda} = - Q(t)   \eqno(6.4a)$$ 
and

$${\dot \rho}_{\rm dm} + 3 H ( 1 + {\omega}_{\rm dm}) \rho_{\rm dm} =
 Q(t)   \eqno(6.4b)$$ 
with $ \rho_{\Lambda}(t)$ and $\rho_{\rm dm}(t)$, being $dark$ $ energy$ density  and $dark$ $matter$ density respectively at cosmic time $t$. Here $H = {\dot a}/a,
{\omega}_{\Lambda} = p_{\Lambda}/\rho_{\Lambda} = - 0.4 $ and ${\omega}_{\rm
  dm} = p_{\rm   dm}/\rho_{\rm dm}$ . $Q(t)$ is the loss (gain) term for
DE(DM) respectively. 

Batchelor \cite{kb67} has pointed out that dissipation is natural for all material fluids,
barring superfluids . So, presence of a dissipative term for dark
matter DM is reasonable. In eq.(6.4b), this term is obtained by taking

$$ Q(t) = 3 n H \rho_{\rm dm} \eqno(6.5a)$$
(with $n$ being the real number) as in \cite{wz} without any harm to
physics. With this setting for time-dependent arbitrary function $Q(t)$, eq.(6.4b)is obtained as

$${\dot \rho}_{\rm dm} + 3 H (1 - n + {\rm w} ) \rho_{\rm dm} = 0 .  \eqno(6.5b)$$
Here $- n \rho_{\rm dm}$ given by $Q(t)$ acts as dissipative pressure
for DM. As $ Q(t)$ is proportional to $\rho_{\rm dm}$, this setting does not
disturb the perfect fluid structure as required by the `cosmological
principle'. Thus eq.(6.5b) yields the effective pressure for DM as

$$ p_{\rm (dm, eff)} = (- n + {\rm w}) \rho_{\rm dm}  .  \eqno(6.5c)$$

Now connecting eqs.(6.4b) and (6.5) and integrating, it is obtained that
$$\rho_{\rm dm} = A a^{3 (n  - {\omega}_{\rm dm} -1)}  \eqno(6.6a)$$
with
$$ A = 0.23 \rho_{{\rm cr},0} a_0^{- 3 (n -1)}      \eqno(6.6b)$$ 
using current value of $dark$ $matter$ density $\rho_{\rm dm,0} = 0.23 \rho_{{\rm cr},0}$ and $a_0 = a (t_0).$

Connecting eqs.(6.4a), (6.5) and (6.6a) and integrating, it is obtained that

$$\rho_{\Lambda} = \frac{D}{a^{3( 1 + {\omega}_{\Lambda})}} - \frac{n}{n +
  {\omega}_{\Lambda} - {\omega}_{\rm dm}} \rho_{\rm dm} \eqno(6.7a)$$ 
where
$$D =  10^{6} a_{\rm ew}^{3( 1 + {\omega}_{\Lambda})} {\rm GeV}^4  \eqno(6.7b)$$ 
using the initial condition $\rho_{\rm dm} = 0$ at $t = 0$.

\smallskip

\noindent {\bf (b) \underline{Expansion of the universe,  when $\rho_{\Lambda} > \rho_{\rm dm}$}}

\smallskip

Friedmann equation is given as

$$\Big(\frac{\dot a}{a} \Big)^2 = \frac{8 \pi G_N}{3} (\rho_{\Lambda} + \rho_{\rm dm} +
\rho_r) \simeq \frac{8 \pi G_N}{3} (\rho_{\Lambda} + \rho_{\rm dm} ) \eqno(6.8)$$
as $(\rho_{\Lambda} + \rho_{\rm dm} )$ dominates over $\rho_r$, which is clear from
eqs.(6.2),(6.6) and (6.7).

In this case, the equation (6.8) reduces to

$$H^2 = \Big(\frac{\dot a}{a} \Big)^2 \simeq \frac{8 \pi G_N D}{3a^{3( 1 +
    {\omega}_{\Lambda})}}  \eqno(6.9)$$
yielding the solution

$$ a(t) = a_{\rm ew}\Big[ 1 + \sqrt{\frac{8 \pi G_N D}{3}} \frac{t}{a_{\rm ew}^{9/10}}\Big]^{10/9} $$

$$= a_{\rm ew}\Big[ 1 + 2.89 \times 10^{-16} t \Big]^{10/9}  \eqno(6.10a)$$
with ${\omega}_{\Lambda} = - 0.4$, $a_{\rm ew} = a(t=0)$ and 
$$ \sqrt{\frac{8 \pi G_N D}{3}} = 2.986 \times 10^{-16} a_{\rm ew}^{9/10} {\rm GeV}\eqno(6.10b)$$
( $G_N = M_P^{-2}, M_P = 10^{19} {\rm GeV}$).

 It shows an accelerated expansion of the universe from the beginning itself as ${\ddot a}/a > 0$. 

Connecting eqs.(6.7) and (6.10)

$$\rho_{\Lambda} = \frac{ 10^{6}}{{\Big[ 1 + \sqrt{\frac{8 \pi G_N D}{3}} \Big(\frac{t}{a_{\rm ew}^{0.9}}\Big) \Big]}^2} - \frac{n}{n +   {\omega}_{\Lambda} - {\omega}_m} \rho_{\rm dm}. \eqno(6.11)$$
But, the present universe obeys the condition

$$\Omega_{\Lambda,0} + \Omega_{{\rm m},0} = 1, \eqno(6.12 a)$$
where $\Omega_{{\rm m},0} = \Omega_{{\rm dm},0} + \Omega_{{\rm r},0}.$ Here $\Omega_0 = \rho_0/\rho_{{\rm cr},0}$ and $ \rho_{{\rm cr},0} = 3H_0^2/{8 \pi G_N}$ ($ H_0$  being Hubble's constant) for the present universe. With these values the present $critical$ $density$ is calculated as

$$ \rho_{{\rm cr},0} = 3H_0^2/{8 \pi G_N} \simeq 1.2 \times 10^{-47} {\rm GeV}^4 \eqno(6.12 b)$$
using $H_0 = h_0/t_0$ with $h_0 = 0.68$  and the present age of the universe $t_0 = 13.7 Gyr = 6.6 \times 10^{41} {\rm GeV}^{-1}$ \cite{abl}.

As, in the present universe, $\rho_m$ is dominated by $cold$ $dark$ $matter$ density, $\omega_m = 0$ is taken in eq.(6.11). Now, conecting eqs.(6.11) and (6.12), it is obtained that

$$ n = 0.47  \eqno(6.13)$$

Using eq.(6.10), the scale factor $a_d$ at the decoupling time $t_d = 1.85
\times 10^{37}{\rm GeV}^{-1} $ and the present scale factor $a_0$ are obtained as

$$ a_d \simeq 1.38 \times 10^{24} a_{\rm ew} \eqno(6.14)$$
and

 $$ a_0 \simeq 1.58 \times 10^{29} a_{\rm ew} \eqno(6.15)$$
using the present age of the universe given above.

The  equation(6.10) shows an accelerated growth of the scale factor when $\rho_{\Lambda} > \rho_{\rm dm}$ exhibiting non-adiabatic expansion of the universe. It implies non-conservation of  entropy of the universe, which is in contrast to the decelerated adiabatic expansion of SMU,  when cosmological model is radiation-dominated or matter-dominated.

In what follows, scale factor dependence of temperature and entropy is obtained in an emperical manner, based on current temperature of the microwave background radiation $T_0 = 2.73 K$,  the initial temperature $ T_{\rm ew} = 78.5{\rm GeV} = 9.1 \times 10^{14}K$ given by eq.(5.9) and $ a_0 \simeq 1.58 \times 10^{34} a_{\rm ew}$ given by eq.(6.15). Thus ,  the required emperical relation is obtained as

$$ \Big[ {T}/{T_{\rm ew}} \Big]^{103/50} = \frac{a_{\rm ew}}{a} \eqno(6.16)$$
which yields the decoupling temperature as 

$$ T_d = T_{\rm ew} \Big[\frac{a_{\rm ew}}{a_d} \Big]^{50/103} = 1740 K, \eqno(6.17)$$
 using eqs.(5.9) and (6.14). This value is much lower than $T_d =3000 K$, obtained in the standard big - bang cosmology. These drastic changes are due to dominance of the $dark$ $energy$ in the proposed model.

Using eq.(5.8b), entropy of the universe is calculated as

$$ S = \frac{7 \pi^2}{90} a^3 T^3 .  \eqno(6.18)$$

Current value of entropy $S_0$ is supposed to be $10^{87}$. ``How could so
high entropy of the universe be generated  ?'' is an old question. A solution
to this problem was suggested in a seminal paper on inflationary model of the
early universe by Guth \cite{ahg} and  modified version of the same by Linde \cite{adl} and Albrecht and Steinhardt \cite{apj}. Here, a different answer to this question is provided on the basis of results obtained above.

Connecting eq.(6.15) and eq.(6.18) and current vlues of entropy as well as temperature $T_0 = 2.73 K$ of the universe, it is obtained that 

$$ a_{\rm ew} = 0.25  \eqno(6.19)$$
as

$$ S_0 = 10^{87} = \frac{7 \pi^2}{90} (1.58 \times 2.73 \times 10^{29}a_{\rm ew})^3$$

Connecting eqs.(6.16) and (6.18) it is obtained that entropy grows with the scale factor $a(t)$ as 

$$ S = \frac{7 \pi^2}{90} T_{\rm ew}^3 a_{\rm ew}^{150/103} a^{159/103} \eqno(6.20 a)$$
with the initial value

$$S_{\rm ew} = 9 \times 10^{42} \eqno(6.20 b)$$

The equation (6.6) yields the rate of production of HDM

$${\dot \rho_{\rm hdm}} = - 5.9 \times 10^{-64} \frac{a_0^{1.59}}{a^{3.48}} \eqno(6.21)$$
using $\omega_{\rm dm} = 1/3$ for $hot$ $dark$ $matter$. This equation shows
that rate of production of HDM increases with growing scale factor. So,
production of HDM is responsible for increasing entropy in the universe. In
SMU, number of photons decide entropy of the universe. But, in PM, photons and
HDM both are responsible for entropy. As energy of HDM increases due to its
production, owing to decay of DE, entropy is generated in this model. It is
unlike SMU, where entropy remains conserved.

Connecting eqs.(6.6) and (6.16), temperature dependence of $\rho_{\rm dm}$ is obtained as

$$ \rho_{\rm dm} = 0.23 \rho_{{\rm cr},0} \frac{a_{\rm ew}^{3 (n  - {\omega}_{\rm dm} -1)}}{a_0^{ 3 (n -1)}} \Big[ {T_{\rm ew}}/{T} \Big]^{6.18(n  - {\omega}_{\rm dm} -1)}. \eqno(6.22)$$

Putting $\omega_{\rm dm} = 1/3 (0)$ for HDM (CDM), in eq.(6.22), ratio of CDM and HDM densities,$\rho_{\rm cdm}$ and $\rho_{\rm hdm}$, is obtained as

 $${\rho_{\rm cdm}}/{\rho_{\rm hdm}} = a_{\rm ew} \Big[{T_{\rm ew}}/{T} \Big]^{2.06} \eqno(6.23)$$
 Eq.(6.23) shows that creation of HDM is higher at high temperature. But as temperature falls down, creation of CDM supersedes the production of HDM ( production of $dark$ $matter$ owing to decay of $dark$ $energy$). For the current universe, the ratio of eq.(6.23) is obtained as

$${\rho_{\rm cdm ,0}}/{\rho_{\rm hdm, 0}} = 2.1 \times 10^{29} \eqno(6.24)$$
using numerical values of $a_{\rm ew}, T_{\rm ew}$ and $T_0$ (given above) in eq.(6.23). It shows that currently, $\rho_{\rm hdm}$ is almost negligible compared to $\rho_{\rm cdm}$.

The scale factor $a_{\rm sd}$ , upto which  $dark$ $energy$ vanishes due to decay, are obtained as

$$ a_{\rm sd} = 6.78 a_0  \eqno(6.25)$$
 connecting eqs.(6.6),(6.7) and (6.10) as well as using ${ \rho_{\Lambda}} = 0$. Here $\omega_{\rm dm} = 0$ is taken as for $t > t_0$, $dark$ $matter$ content is expected to be dominated by CDM. 

The ratio of densities of $dark$ $energy$, $\rho_{\Lambda}$, and $dark$ $matter$, $\rho{\rm dm}$, is obtained as

$$ {\rho_{\Lambda}}/{\rho_{\rm dm}} = 10 \Big(a_0/a \Big)^{0.21} - \frac{47}{7}. \eqno(6.26)$$
It is interesting to note, from eq.(6.26), that the gap between  $dark$ $energy$ density and  $dark$ $matter$ density decreases with the growing  scale factor.
Connecting eqs.(6.6) and (6.7) and using $ n $ from eq.(6.13). This equation
    shows that  $\rho_{\rm dm}/\rho_{\Lambda} < 1$ for

$$ a_{\rm ew}< a < 3.44 a_0 \quad {\rm and} \quad 0 < t < 3.03 t_0. \eqno(6.27 a,b)$$
The result, given by  eq.(6.27b), solves the $cosmic$ $coincidence$ $problem$ [24]. This approach, for solution of $coincidence$ $problem$  has an advantange that no scalar field is required to represent the $dark$ $energy$, rather it is the contribution of higher-dimensional higher-derivative gravity to the observable universe. It is unlike the case of work in refs.[25-28], where assumed $quintessence$ scalars are used.

\smallskip

\noindent {\bf (c) \underline{The universe,in case $\rho_{\Lambda} < \rho_{\rm dm}$}}

\smallskip

Like eq.(6.27), employing the same procedure, it is also possible
to find that ${\rho_{\rm dm}}/{\rho_{\Lambda}} \ge 1$ for

$$  a \ge 3.44 a_0 \quad {\rm and} \quad  t \ge 3.03 t_0. \eqno(6.28 a,b)$$

So, for $ t \ge 3.03 t_0$, the  the Friedmann equation is written as

$$\Big(\frac{\dot a}{a} \Big)^2 = \frac{8 \pi G_N}{3} (\rho_{\Lambda} + \rho_{\rm dm} + \rho_r) \simeq \frac{8 \pi G_N}{3} {\rho_{\rm dm}} \simeq 0.27 H_0^2 \Big(a_0/a \Big)^{1.74} \eqno(6.29)$$
using eqs.(6.6),(6.8), (6.13) and definition of $\rho_{\rm cr,0}$ (given above). Eq.(6.29) yields the solution 

$$ a(t) = \Big[ a_{\Lambda < {\rm dm}}^{87/100} + 0.43\times 10^{-43} (t - t_{\Lambda < {\rm dm}}) \Big]^{100/87}, \eqno(6.30)$$
where $a_{\Lambda < {\rm dm}} = 3.44 a_0$ and $t_{\Lambda < {\rm dm}} = 3.03 t_0$. Eq.(6.30) also shows  an accelerated growth of the scale factor, even though $\rho_{\Lambda} < \rho_{\rm dm}$. It is interesting to see that this expansion is faster than the expansion in the interval $0 < t < 3.03 t_0$.

From eq.(6.25), the scale factor upto which $dark$ $energy$ vanishes is $a_{\rm sd} = 430 a_0$. So, eq.(6.30) yields the corresponding time  as

$$ t_{\rm sd} = 3.7 t_0 = 50.69 {\rm Gyr}. \eqno(6.31)$$

It shows that decay of $dark$ $energy$ will continue in the interval $3.03 t_0 < t < 3.7 t_0$ also. So, during this time interval,  the $dark$ $matter$ will follow the rule, given by eq.(6.6).

Using eqs.(6.16),(6.18), (6.25)and (6.31), $T_{\rm sd}$ and $S_{\rm sd}$ are calculated as

$$ T_{\rm sd} = T_0 \Big( \frac{a_0}{a_{\rm sd}} \Big)^{50/103} = 0.395 T_0 = 1.08 K  \eqno(6.32 )$$

$Dark$ $matter$ density at $t = t_{\rm sd}$ is obtained as  

$$\rho_{\rm dm(sd)} = 1.32 \times 10^{- 49} {\rm GeV}^4 \eqno(6.33)$$
using $a_0$ and $a_{\rm sd}$ in eq.(6.6).

As $dark$ $energy$ vanishes when $t \ge t_{\rm sd}$, content of the universe will be dominated by CDM ,which is   pressureless non-baryonic matter obeying the conservation equation

$$ {\dot \rho_{\rm dm}} + 3 H \rho_{\rm dm} = 0.$$
This equation yields scale factor dependence of $\rho_{\rm dm}$ as

$$\rho_{\rm dm} = \rho_{\rm dm(sd)} \Big( a_{\rm sd}/ a(t) \Big)^3 = \frac{2.54 \times 10^{39}}{a^3(t)} \eqno(6.34)$$
for $ t > 3.7 t_0.$ 

Beyond the age of the universe $3.7 t_0$, the Friedmann equation looks like

$$\Big(\frac{\dot a}{a} \Big)^2 = \frac{8 \pi G_N}{3} \rho_{\rm dm} =  \frac{212.8}{a^3(t)} \eqno(6.35)$$
using eq.(6.34).

Eq.(6.35) yields the solution

$$ a(t) = \Big[ a_{ {\rm sd}}^{3/2} \pm 14.6 (t - t_{ {\rm sd}}) \Big]^{2/3}, \eqno(6.36)$$
with $a_{\rm sd}$ and $t_{\rm sd}$ given by eqs.(6.25) and (6.31) respectively.

On taking (+) sign in eq.(6.36), decelerated expansion of the universe is obtained beyond $t > 3.7 t_0$ continuing for ever. But the (-) sign, in eq.(6.36), exhibits a contracting universe beyond $t > 3.7 t_0$. So, ultimately the contracting universe is expected to collapse to a very small size with scale factor, possibly equal to $a_{\rm ew} = 0.25$. From eq.(6.36), with (-) sign, the collapse time is calculated as

$$ t_{\rm col} = t_{\rm sd} + \frac{a^{3/2}}{14.6} = 18.08 t_0 = 247.73 {\rm
  Gyr} \eqno(6.37)$$
using $a_{\rm sd} = 6.78 a_0$ and $t_{\rm sd} = 3.7 t_0$.

\bigskip

\centerline{\bf 7. Elementary particles, Primordial nucleosynthesis and}

 \centerline{\bf Structure formation  }

\smallskip

\noindent \underline{7.1 Creation of particles}

\smallskip

In SMU, it is assumed that elementary particles such as leptons, mesons,
nucleons and their anti-particles were produced at the epoch of
big-bang. Here, production of scalar and spin-1/2 particles is proposed in the very early
universe due to topological changes caused by expansion of the universe.

A lot of work has been done in the past,where production of particles are
discussed in curved space-time. This mechanism is based on the fact that, in
the flat space-time, vacuum state in the Fock space is stable, if there is no
source term in the action of the matter field. In curved space-time, vacuum is
unstable even in the absence of the source term in the action. It turns out
that gravity makes the initial vacuum state  $|0, {\rm in}>$ unstable
such that $|0, {\rm in}> \ne |0, {\rm out}>,$ where $|0, {\rm out}>$ is the
other vacuum state in the new Fock space. The instability of vacuum state
shows production of particles where conformal symmetry is broken. Moreover,
changing gravitational field contributes rest mass to produced particles
\cite{sk96, ndb, em}. Here $|0, {\rm in}>$ state is defined when $t \to 0$ and
$|0, {\rm out}>$ is obtained for $t \to \infty$.

\smallskip

\noindent \underline{(a) Creation of spinless particles}

\smallskip

The scalar field $\phi$ obeys the Klein-Gordon equation

$$ ({\Box} + m_{\phi}^2 ) \phi = 0,    \eqno(7.1)$$
where $m_{\phi}$ is mass of $\phi$. Expanding $\phi$ in terms of mode $k$,
eq.(7.1) is written as

$$( 1 + 2.89 \times 10^{-16} t ) {\ddot \phi_k} + \frac{28.9}{3} \times 10^{-16}{\dot \phi} + \Big[m_{\phi_k}^2 (1 + 2.89 \times 10^{-16} t ) $$
$$- \frac{k^2}{a^2_{\rm    ew}} ( 1 + 9.86 \times 10^{-16} t )^{-11/9} \Big]
\phi_k = 0  \eqno(7.3)$$
using the scale factor $a(t)$, given by eq.(6.10a).

For small $t$, eq.(7.3) is obtained as

$$ \tau \frac{d^2 \phi^{\rm in}_{ k}}{d \tau^2} + \frac{10}{3} \frac{d
  \phi^{\rm in}_{ k}}{d \tau} +  10^{31} \Big[- \frac{20
  k^2}{9 a^2_{\rm  ew}} + \Big( m^2_{\phi} + \frac{11 k^2}{9 a^2_{\rm  ew}}
  \Big) \tau \Big] \phi^{\rm in}_{ k} = 0 , \eqno(7.4a)$$
where
$$ \tau =  1 + 2.89 \times 10^{-16} t . \eqno(7.4b)$$

Eq.(7.4) yields the normalized solution

$$\phi^{\rm in}_{ k} = [2 (2 \pi)^3 \sqrt{b_2} ]^{-1/2} e^{- i \tau
  \sqrt{b_2}} _1F_1 \Big(\frac{5}{3} + i \frac{b_1}{2 \sqrt{b_2}} , \frac{10}{3},  2 i \tau \sqrt{b_2} \Big) $$

$$ \approx [2 (2 \pi)^3 \sqrt{b_2} ]^{-1/2} e^{- i \tau
  \sqrt{b_2}} \Big[ 1 + \frac{( - 3 b_1 + 10 i \sqrt{b_2}) \tau}{10} \Big],
 \eqno(7.5a)$$
where
$$ b_1 = - \frac{20}{9} \frac{( 10^{15.5} k)^2}{ a^2_{\rm  ew}} \eqno(7.5b)$$
and

$$ b_2 = 10^{30}  \Big( m^2_{\phi} + \frac{11 k^2}{9 a^2_{\rm
    ew}}  \Big) . \eqno(7.5c)$$
Here $_1F_1 (a,b,c)$ is the confluent hypergeometric function.

For large $t$, eq.(7.3) reduces to

$$ \tau \frac{d^2 \phi^{\rm out}_{ k}}{d \tau^2} + \frac{10}{3} \frac{d
  \phi^{\rm out}_{ k}}{d \tau} + ( 10^{15.5} m_{\phi})^2 \tau
  \phi^{\rm out}_{ k}  = 0 , \eqno(7.6)$$
which integrates to

$$ \phi^{\rm out}_{ k} = \tau^{-7/6} \Big[ c_1 J_{-7/6} (A \tau) + c_2
Y_{-7/6} (A \tau) \Big] , \eqno(7.7a)$$

where 

$$ A = 10^{15.5} m_{\phi} . \eqno(7.7b)$$
and $J_p(x)$ and $Y_p(x)$ are Bessel's function of first and second kind
respectively. For large $x$, Bessel's functions can be approximated as 
$$ J_p(x) \simeq \frac{cos (x - \pi/4 - p \pi/2)}{\sqrt{\pi x/2}} \eqno(7.7c)$$
and
$$ Y_p(x) \simeq \frac{sin (x - \pi/4 - p \pi/2)}{\sqrt{\pi x/2}}. \eqno(7.7d)$$

Using these approximations, when $\tau$ is large, eq.(7.7a) looks like

$$ \phi^{\rm out}_{ k} \simeq \frac{\tau^{-5/3}}{\sqrt{\pi A/2}} [c_1 cos (A
\tau + \frac{\pi}{3}) + c_2 sin (A \tau + \frac{\pi}{3}) ] \eqno(7.8)$$

Solutions (7.5a) and (7.8) yield number of produced spinless particles (for
mode $k$) per unit volume as

$$ |\beta_k|^2 = \frac{\tau^{-10/3}}{\pi^2 A^2 b_2} |X|^2 \ne 0 , \eqno(7.9a)$$
where
\begin{eqnarray*}
 X &=& \Big[ \Big(- \frac{3 b_1 \sqrt{b_2} \tau}{10} - \frac{2}{3} \sqrt{b_2}
\Big) cos(A \tau) - A\sqrt{b_2} \tau sin (A \tau) \Big]  + i \Big[
\Big\{\sqrt{b_2} \tau \\&& - \frac{3 b_1}{10} + \frac{5}{3} \Big(1 -
\frac{3 b_1}{10} \Big) \tau^{-1} \Big\} cos(A \tau) + \Big(1 -
\frac{3 b_1}{10} \Big) A sin(A \tau) \Big].
\end{eqnarray*}
$$ \eqno(7.9b)$$
$\beta_k$ is defined in Appendix B. Non-zero $ |\beta_k|^2$ shows creation of
spinless particles.

\smallskip

\noindent \underline{(b) Creation of spin-1/2 particles}

\smallskip

The spin-1/2 field $\psi$ satisfies the Dirac equation

$$ (i \gamma^{\mu} D_{\mu} - m_f ) \psi = 0 , \eqno(7.10)$$ 
where $m_f$ is mass of $\psi$, $\gamma^{\mu}$ are Dirac matrices in curved space-time and $ D_{\mu}$
are covariant derivatives defind in Appendix B. $\psi$ can be written as

$$ \psi = {\sum_{s = \pm 1}} {\sum_{k}} (b_{k,s} \psi_{I k,s} + d^{\dagger}_{k,s}
\psi_{II k,s})  \eqno(7.11)$$
with $b_{k,s}$ and $d_{k,s}$, given in Appendix B. Eqs.(7.10) and (7.11) yield

$$(i \gamma^{\mu} D_{\mu} - m_f )\psi_{I k,s}  = 0 , \eqno(7.12a)$$ 

$$(i \gamma^{\mu} D_{\mu} - m_f )\psi_{II k,s}  = 0 . \eqno(7.12b)$$ 

Now using the operator $(- i \gamma^{\mu} D_{\mu} - m_f )$ from left of
eqs.(7.12a) and (7.12b), it is obtained that

$$ ({\Box} + \frac{1}{4} R + m_f^2 ) {\tilde \psi} = 0 , \eqno(7.13)$$
where ${\tilde \psi} = \psi_{I k,s} (\psi_{II k,s}).$

Writing
$$ \psi_{I k,s} = f_{I k,s}(t) e^{i{\vec k}.{\vec x}}u_s$$
and
$$ \psi_{II k,s} = f_{II k,s}(t) e^{i{\vec k}.{\vec x}}{\hat u_s},$$
eq.(7.13) is obtained as

$$ \tau \frac{d^2 {\tilde f}}{d \tau^2} + \frac{10}{3} \frac{d {\tilde f}}{d
  \tau} +  10^{30}  \Big[m_f^2\tau -
  \frac{k^2}{\tau^{11/9}} - \frac{4.93 \times 10^{-15}}{3 \tau} \Big]{\tilde f} 
    = 0 , \eqno(7.14)$$
where ${\tilde f} = f_{I k,s} (f_{II k,s}).$

For small t , eq.(7.14) looks like

$$ \tau \frac{d^2 {\tilde f}}{d \tau^2} + \frac{10}{3} \frac{d {\tilde f}}{d
  \tau} +  10^{30}  \Big[(m_f^2 + \frac{11 k^2}{9 a_{\rm ew}^2})
  \tau - \frac{20 k^2}{9 a_{\rm ew}^2}\Big] {\tilde f} = 0 . \eqno(7.15)$$

This equation is like the equation (7.4a). So, its solution has the same
form. Using this solution in $\psi_{I k,s}$ defined above , it is obtained that
$$ \psi^{\rm in}_{I k,s} = \frac{e^{-i \tau \sqrt{b_2}}}{\sqrt{2 (2 \pi)^3
      \sqrt{b_2}}} \Big[ 1 + \frac{( - 3 b_1 + 10 i \sqrt{b_2})}{10} \tau
      \Big] e^{i {\vec k}.{\vec x}}u_s . \eqno(7.16)$$

For large t , eq.(7.14) reduces to

$$ \tau \frac{d^2 {\tilde f}}{d \tau^2} + \frac{10}{3} \frac{d {\tilde f}}{d
  \tau} +  10^{30} m_f^2 \tau {\tilde f} = 0  \eqno(7.17)$$
yielding

$$ \psi^{\rm out}_{II k,s} = \frac{ \tau^{-5/3}}{\pi^2 A} cos (A \tau)
e^{i{\vec k}.{\vec x}}{\hat u_s}. \eqno(7.18)$$

Using $\psi^{\rm in}_{I k,s}$ and $ \psi^{\rm out}_{II k,s}$ in the definition
of $\beta_{k,s}$, given in the Appendix B,

$$\beta_{k,s} = (2 \pi)^3 \tau^{-5/3} \frac{cos (A \tau)}{\pi^2 A}\frac{e^{-i \tau \sqrt{b_2}}}{\sqrt{2 (2 \pi)^3
      \sqrt{b_2}}} \Big[ 1 + \frac{( - 3 b_1 + 10 i \sqrt{b_2})}{10} \tau
      \Big] . \eqno(7.19)$$

So, the number of created spin-1/2 particles per unit volume is obtained from
eq.(7.19)as

$$ |\beta_{k,s}|^2 = \frac{(2 \pi)^2}{2 \pi^2 A \sqrt{b_2}} \tau^{-10/3} \Big[
\Big( 1 - \frac{3 b_1}{10} \tau \Big)^2 + b_2 \tau^2 \Big]. \eqno(7.20)$$

Eqs.(7.9a,b) and (7.20) show creation of spinless and spin-1/2 particles due to
changing gravitational field in the proposed speeded-up model. The scale
factor, given by eq.(6.10a) show that the significant change in the
gravitational field is possible , in this model, when

$$ t > 3.46 \times 10^{15.5} {\rm GeV}^{-1} = 7.2 \times 10^{-9} {\rm sec.}  \eqno(7.21)$$
as $a(t)$ remains almost constant upto this epoch. So, here, particle
production is expected, when the universe is around $2.3 \times 10^{-9} {\rm
  sec.}$ old. Also, these results show that particle production falls down
rapidly as time increases. Thus, like other models, here also particle
production is expected in the early stages of the universe. 

\smallskip

\noindent \underline{7.2 Primordial Nucleosynthesis}

\smallskip

Hydrogen ($H$) is the major component of baryonic matter in the universe. The
next main component is Helium-4 ($^4He$) . Occurrence of other light elements
and metals is very small. It is found  unlikely that abundance of $^4He$,
Deuterium ($D$) and other light elements, in the universe, being caused by burn out of $H$ in stars
\cite{sw, ek} So, it is argued that the required amount of these might have produced in
the early universe.

At the epoch of its formation, helium production depends upon neutron ($n$)
concentration, which is determined by weak interaction reactions given as

$$ n + \nu \leftrightarrows p + e^{-} , n + e^{+} \leftrightarrows p + {\bar
  \nu}  \eqno(7.22)$$
where $p$ stands for proton and $\nu$ for neutrino.
This chemical equillibrium is maintained till weak reaction rate $\Gamma_{\rm
  w} >> H $, where $H$ is the expansion rate of the universe given by eq.(6.9)
  and $\Gamma_{\rm   w} \simeq 1.3 G_F^2 T^5 $ with Fermi constant $G_F = \pi
  \alpha_{\rm   w}/{\sqrt{2} M^2_{\rm   w}} = 1.17 \times 10^{-5} {\rm
  GeV}^{-2}$. With the expansion of the universe, temperature decreases so
  weak interaction rate slows down . As a result, at the freeze-out
  temperature $T_*$,
$$\Gamma_{\rm   w} \simeq H.   \eqno(7.23a)$$ 

Connecting eqs.(6.7b), (6.9), (6.16) and (7.23a), it is obtained that

$$ 1.3 \times (1.17)^2 \times 10^{-10} T_*^5 = 2.89 \times 10^{-16}
\Big(\frac{T_*}{T_{\rm ew}} \Big)^{1.9}, $$
which yields
$$ T_* = 0.9 {\rm MeV} . \eqno(7.23b)$$ 

Eqs.(6.10a,b) and (6.16) yield time dependence of temperature , for this
model, as 

$$ t_{\rm sec} = \frac{4.5}{ T^{1.9}_{\rm MeV}} . \eqno(7.24)$$ 

Using this result, the freeze-out time $t_*$, corresponding to $T_*$, is
obtained as
$$ t_* \simeq 5.55 sec. \eqno(7.25)$$

In SMU,$T_* \simeq 0.86 {\rm MeV}$ and $ t_* \simeq 1 sec$ \cite{muk}. It is
obtained that, in PM, $T_*$ is a bit higher and freeze-out takes place
later. The reason for these differences is the basic difference in developement
of these models. SMU is driven by elementary particles present in the early
universe and it expands adiabatically. The proposed model is driven by dark
energy, with very high density at the beginning and it expands with
acceleration.

Like \cite{re}, in this subsection, temperature $T_9$ is measured in units
$10^9 K$, so $1 {\rm MeV}$ becomes $11.6$ in $T_9$ temperature. Now, eq.(7.24)
looks as

$$ t = \frac{474}{ T_9^{1.9}}  .\eqno(7.26)$$

For large temperature $T_9 > 10,$ neutron abundance is given by

$$ X_n = \Big[ 1 + e^{Q_9/T_9} \Big]^{-1},  \eqno(7.27a)$$
where $Q_9 = m_n - m_p = 15.$ At freeze-out temperature $T_* \simeq 0.9 {\rm
  MeV},$ which is equivalent to $T_9 = 10.44$,

$$ X_n^* = 0.19. \eqno(7.27b)$$

Due to slightly higher freeze-out temperature, in the present model, neutron
concentration is obtained higher than $X_n^* = 0.16$ in SMU.

When $\Gamma_{\rm   w} << H $ i.e. when chemical equillibrium between
$n$ and $p$ frezze-out, neutron concentration is determined through
neutron-decay
$$ n \to p + e^{-} + {\bar \nu}.   \eqno(7.28)$$

Thus, for $t > t_*,$ neutron abundance is given as 

$$ X_n(t) = 0.19 e^{-t/\tau}, \eqno(7.29)$$
where $\tau = (885.7 \pm 0.8) {\rm sec}$. is the neutron life-time.

First step, in the formation of complex nuclei, is 
$$ p + n \leftrightarrow D + \gamma$$
showing chemical equillibrium between deutron $(D)$ , nucleons and photon $(\gamma)$. This
equillibrium gives deutron abundance by weight \cite{ek} as

$$ X_D = 0.2 \times 10^{-12} (\Omega_B h^2) X_n X_p T_9^{3/2} exp(B_D/T_9), \eqno(7.30)$$
where $X_D = 2 n_D/n_B, X_n =  n_n/n_B, X_p = n_p/n_B, X_n + X_p = 1,$
deuterium binding energy $B_D = m_p + m_n - m_D = 2.23 {\rm MeV}$ which yields
$B_{D9} = 25.82 $ (in units of $T_9$) and $(\Omega_B h^2)$ is the baryon
number density.

Light heavy elements $^4He$ is formed through  reactions

$$ n + p \rightarrow D + \gamma,    \eqno(7.31a)$$

$$ (i)D + D \rightarrow ^3He + n, (ii) D + D \rightarrow ^3H + p
\eqno(7.31b,c)$$
and
$$ (ii) D + ^3H \rightarrow ^4He + n, (ii) ^3He + D \rightarrow ^4He + p,
\eqno(7.31d,e)$$
where $^3H$ stands for tritium.

These reactions show that sufficient deuterium abundance is required for
nucleosynthesis of $^4He$, which is supposed to be around $25\%$ of the
baryonic content of the universe. But after weak-interaction freeze out,
deuterium abundance is not sufficient unless temperature is very low. For
example, at $T_9 = 5.8 ( T = 0.5 {\rm MeV}), X_D = 2\times 10^{-12}$ is
obtained using eq.(7.30) and putting $\Omega_B h^2 = 0.05$.

Rates for reactions (7.30b,c) are given as \cite[Appendix]{re}
\begin{eqnarray*}
\lambda_{DD(i)} &=& 3.97 \times 10^5 T_9^{-2/3} e^{-4.258/T_9^{1/3}}[1 + 0.098
T_9^{1/3} + 0.876 T_9^{2/3}
\\&& + 0.6 T_9 - 0.041 T_9^{4/3} - 0.071
T_9^{5/3}] (\Omega_B h^2) {\rm sec}^{-1}\\ \lambda_{DD(ii)} &=& 4.17 \times 10^5 T_9^{-2/3} e^{-4.258/T_9^{1/3}}[1 + 0.098 T_9^{1/3} + 0.518 T_9^{2/3} \\&& - 0.355 T_9 - 0.010 T_9^{4/3} -
0.018T_9^{5/3}] (\Omega_B h^2){\rm sec}^{-1}
\end{eqnarray*}
$$ \eqno(7.32a,b)$$
Making approximations in eqs.(7.30a,b), it is obtained that
\begin{eqnarray*}
\lambda_{DD} &=& \lambda_{DD(i)} + \lambda_{DD(ii)}
\\&& \sim 8.14\times 10^5 T_9^{-2/3} e^{-4.258/T_9^{1/3}}(\Omega_B h^2){\rm
  sec}^{-1} 
\end{eqnarray*}
$$ \eqno(7.33)$$

Mukhanov \cite{muk} has obtained the condition for conversion of sufficient
amount of $D$ to $^3He$ and $^3H$ as

$$ \frac{1}{2} \lambda_{DD} X_D t \sim 1  \eqno(7.34)$$
From eqs.(7.27), (7.30)-(7.33), it is obtained that
$$ 6.43 \times 10^{-6} T_{9i}^{-1.3} e^{\Big(\frac{25.82}{T_{9i}} -
  \frac{4.258}{(T_{9i})^{1/3}} \Big)}(\Omega_B h^2)^2  \sim 1  \eqno(7.35)$$
showing a relation between $\Omega_B h^2$ and $T_{9i}$. $T_{9i}$ is the
  temperature at which nucleosynthesis takes place.
For $\Omega_B h^2 = 0.05$, this equation yields
$$ T_{9i} \sim  1.16, \eqno(7.36)$$
which corresponds to $ T_i = 0.1 {\rm MeV}$. It shows that if baryon number
density $\Omega_B h^2$ is $\sim 0.05,$ nucleosynthesis of $^4He$ may take place
at temperature $ T_i = 0.1{\rm MeV}$. In SMU, this temperature is $\sim
0.086{\rm MeV}$.

Eqs.(7.29) and (7.35) yield deuterium concencentration at $T_i$ as
$$X_{Di} \simeq 8.93 \times 10^{-6}    \eqno(7.37)$$

It happens at the cosmic time
$$ t_i = 357.5 {\rm sec} . \eqno(7.38)$$
Thus, it is obtained that, in the present model, nucleosynthesis of $^4He$
begins much later compared to SMU, where it happens at time $\sim 100{\rm
  sec}$ \cite{muk}. But, it is still less than life-time of a neutron, which
is required for the necleosynthesis.

The final $^4He$ abundance is determined by available free neutrons at cosmic
time $t_i$. As total weight of $^4He$ is due to neutron and proton, so its
final abundance by weight is given by \cite{re} as
$$ X^f_{\alpha} = 2 X_n^* exp(-t^f/\tau ) =  2 X_n^* exp(- 0.535/(T^f_{9})^{1.9} \eqno(7.39)$$
using eqs.(7.26) and $\tau = 886.5 {\rm sec}$. Here $\alpha$ stands for $^4He$.

The rate of production of $^4He$ is given by the equation
$$ {\dot X_{\alpha}} = 2 \lambda_{DD(ii)} X_D^2 . \eqno(7.40)$$
As free neutrons are captured in $^4He$, helium production dominates neutron
decay. It happens when \cite{re}

$$ 2{\dot X_{\alpha}} = \frac{X_n}{\tau}. \eqno(7.41)$$
Approximating $\lambda_{DD(ii)},$ given by eq.(7.32b), around $T_9 = 1.16,$ it
is obtained that

$$ \lambda_{DD(ii)} \simeq 6.98 \times 10^5 T_9^{-2/3} exp(- 4.258/T_9^{1/3})
(\Omega_B h^2). \eqno(7.42)$$

Eqs.(7.29), (7.38)-(7.40) yield

$$\frac{1}{2 \tau} \simeq 6.98 \times 10^5 T_9^{7/3} e^{(\frac{51.64}{T_9} -
  \frac{4.258}{T_9^{1/3}})} \times 0.19 e^{(- \frac{0.535}{T_9^{2.133}})} $$
$$ \times (0.2 \times 0.81 \times 10^{-12})^2 (\Omega_B h^2)^3,$$
which implies that

$$ T^f_9 = \frac{51.64}{37.96 - \frac{7}{3} ln T_9 - 3 ln(\Omega_B h^2) +
  \frac{4.258}{T_9^{1/3}} + \frac{0.535}{T_9^{2.133}} }. \eqno(7.43)$$

At $T = T_{9i} = 1.16,$

$$ T^f_9 = \frac{51.64}{42.08 - 3 ln(\Omega_B h^2)}. \eqno(7.44)$$

Using eq.(7.44) in eq.(7.39), helium abundance by weight, as function of baryon
number density $(\Omega_B h^2)$,is obtained as

$$ X^f_{\alpha} \simeq 0.38  exp \Big[ - 0.535\Big \{\frac{42.08 - 3 ln(\Omega_B h^2)}{51.64}\Big \}^{1.9} .. \eqno(7.45)$$

For $\Omega_B h^2 = 0.05,$ it is obtained that
$$ X^f_{\alpha} \simeq 0.23 . \eqno(7.46)$$

For $T < T^f_9,$ neutron abundance is governed by ${\dot X_n} = - 2 {\dot
  X_{\alpha}}$ with ${\dot   X_{\alpha}}$ given by eq.(7.39). So,

$$ {\dot X_n} = - 2 \lambda_{DD(ii)} X_D^2, $$ 
where $X_D$ and $\lambda_{DD(ii)}$ are given by eqs.(7.30) and (7.41)
respectively. Using eq.(7.26) and integrating, this equation yields neutron
abundance below $ T^f_9$ as

$$ X_n = \Big[(1/X_n^{\alpha}) + (\Omega_B h^2)^3 e^{-37.62} T_9^{0.43} \Big(
e^{51.64/T_9} - e^{51.64/T^f_9} \Big) \Big]^{-1}, \eqno(7.47a)$$
which shows that concentration of free neutrons drops very fast. It is due to
capture of free neutrons in $^4 He$. From eq.(7.29), it is obtained that
neutron and deuterium concentrations become equal at temperature $T^d_9$ given
as \cite{re}
$$ T^d_9 = \frac{25.82}{29.45 - ln(\Omega_B h^2) - \frac{3}{2} ln T^d_9 }.\eqno(7.47b)$$ 
Deutron concentration is given by \cite{re}

$$ {\dot X_D} \simeq - 2\lambda_{DD(ii)} X_D^2 = - 2\times 8.14\times 10^5
T^{-2/3} (\Omega_B h^2) e^{-4.258/T_9^{1/3}} X_D^2, $$
which integrates to
\begin{eqnarray*}
 X_D &=& \Big[\Big(1/X_D(T^d_9)\Big) + (\Omega_B h^2) e^{20.16}
\{(T^d_9)^{-2.57} e^{-4.258/(T^d_9)^{1/3}} \\&&- (T_9)^{-2.8}
e^{-4.258/(T_9)^{1/3}}\}\Big]^{-1},
\end{eqnarray*}
$$ \eqno(7.47c)$$
where $X_D(T^d_9) = X_n(T^d_9) .$

Eqs.(7.47a) and (7.47c) show that, below $T^f_9$, abundance of free neutron
and deuterium drop rapidly in PM also like SMU. 
 
\smallskip

\noindent \underline{7.3 Growth of inhomogeneities in the proposed model}
\smallskip

For small inhomogeneities, linear perturbation of Einstein equations is
enough. In this case, contrast density $\delta = \delta \rho_{\rm
  dm}/\rho_{\rm dm}$ (with $\rho_{\rm dm}$, being the cold dark matter energy
density and $\delta\rho_{\rm dm}$ is small fluctuation in $\rho_{\rm dm}$
presenting inhomogeneity), obeys a linear differential equation in the
homogeneous model of the universe \cite{pd}. Due to linearity, $\delta$ can be expanded
in modes $k$ as 

$$ \delta = {\sum_k} \delta_k e^{i {\vec k}.{\vec x}}.    \eqno(7.48)$$
For modes $k$ with proper wavelength $\digamma < H^{-1}(t) (H^{-1}(t) $ being
the Hubble's radius), the perturbation equation looks like 
$$ {\ddot \delta_k} + 2 \frac{\dot a}{a} {\dot \delta_k} + \Big( \frac{k^2
  v_s^2}{a^2} - 4 \pi G \rho_{\rm dm} \Big) \delta_k = 0,  \eqno(7.49)$$
which is free from gauge ambiguities \cite[eq.(4.158)]{pd}. So, $\delta_k$
  depends on $t$ only.

Eqs.(6.5c) and (6.13) give effective pressure for DM as 
$$ p_{\rm (dm, eff)} = (- 0.47 + {\rm w_{dm}}) \rho_{\rm dm}. \eqno(7.50)$$

It yields
$$ v_s^2 = \frac{d  p_{\rm (dm, eff)}}{d \rho_{\rm dm}} = - 0.47 + {\rm w}
.\eqno(7.51)$$

Connecting eqs.(6.6), (6.13), (7.49) and (7.51), the differential equation for
$\delta_k$ is obtained as
$${\ddot \delta_k} + 2 \frac{\dot a}{a} {\dot \delta_k} + \Big[(- 0.47 + {\rm
  w_{dm}}) \frac{k^2}{a^2} - \Big(0.345 H_0^2 a_0^{1.59}/a^{3(0.53 + {\rm
  w_{dm}} )} \Big)\Big] \delta_k = 0 \eqno(7.52)$$
with $\rho_{\rm cr,0} = 3 H^2_0/{8 \pi G}$.

Using eq.(6.10), eq.(7.52) looks like
$$ \tau \frac{d^2 \delta_k}{d \tau^2} + \frac{10}{3} \frac{d \delta_k
}{d \tau} +  10^{31} \Big[(- 0.47 + {\rm
  w_{dm}}) \frac{k^2}{\tau^{11/9}}$$
$$ - \frac{0.345 H_0^2 a_0^{1.59}}{\tau^{10(0.23 +
  {\rm   w_{dm}} )/3}} \Big] \delta_k = 0 ,\ eqno(7.53)$$
where $\tau$ is defined in eq.(7.4b).

As mentioned above, for CDM $\rm {w_{dm}} = 0$, so eq.(7.52) is approximated as 
$$ \tau \frac{d^2 \delta_k}{d \tau^2} + \frac{10}{3} \frac{d \delta_k
  }{d \tau}  - \frac{3.45\times 10^{29} H_0^2 a_0^{1.59}}{\tau^{2.3/3}} \Big] \delta_k = 0 , \eqno(7.54)$$
which is integrated to
\begin{eqnarray*}
\delta_k &=& \tau^{-11/18} \Big[c_1 H^{(1)}_{110/21} (2 i b \tau^{7/60}) + c_2
H^{(2)}_{-110/21} (2 i b \tau^{7/60}) \Big] \\
& =&  \tau^{-7/6} \Big[(c_1+ c_2) J_{110/21} (2i b \tau^{7/60}) + i (c_1 - c_2)
Y_{110/21} (2 i b \tau^{7/60}) \Big], 
\end{eqnarray*}
 $$ \eqno(7.55)$$
where $i = \sqrt{-1}$ and  $b^2 = 3.45 \times 10^{30} H_0^2 a_0^{1.59}.$  Here $H^{(1)}_p(x)$ and
$H^{(2)}_p(x)$ are Hankel's functions as well as $J_p(x)$ and $Y_p(x)$ are
Bessel's functions. Moreover $c_1$ and $c_2$ are integration constants.

As discussed in section 6, CDM dominates HDM in the late universe, so
structure formation is expected for large $\tau$. Using approximations
(7.7c,d)for Bessel functions, for large $\tau$ in eq.(7.55)and doing some manipulations,
it is obtained that

$$\delta_k \approx \tau^{-0.67} e^{b \tau^{7/60}} , \eqno(7.56)$$
which shows growth of structure formation as cosmic time $t$ increases.

\bigskip

\centerline{\bf 8. Summary of Results  }

\smallskip

In contrast to the original $big-bang$ theory, the proposed
cosmology answers many basic questions as (i)``What is the $fireball$?'', (ii)
``How does it burst out?'', (iii) ``What is the background radiation?'' and
``What are initial values of temperature and energy density?'' Moreover, the
proposed cosmological model is free from initial singularity. It is found that
the $dark$ $energy$ violates the $strong$ $energy$ $condition$ showing
``bounce'' of the universe, which is consistent with $singularity - free$
model of cosmology \cite{sk93,  sk98, sk99}. The initial scale factor is computed to be $a_{\rm ew} = 0.25$. It is in contrast to the standard model of cosmology SMU, which encounters with singularity having zero scale factor, infinite energy density and infinite temperature.

The present value of $dark$ $energy$ density is supposed to be $\sim 7.3
\times 10^{-48} {\rm Gev}^4$, which is 53 orders below its initial
values $10^6 {\rm Gev}^4$ . The question ``How does $dark$ $energy$ falls by 53 orders in the
current universe?'' is answered by the result, in section 6(a), showing that
$dark$ $energy$ decays to $dark$ $matter$, obeying the rule, given by
eq.(6.7). Upto the decoupling time $t_d \simeq 386 kyr$, matter remains in
thermal equllibrium with the background radiation, so produced $dark$ $matter$
upto $t_d$ is supposed to be HDM. But when $t >t_d$, production of CDM is more
than HDM. The current ratio of HDM density and CDM density is found to be
$5 \times 10^{-30}$. In section 6, it is found that creation of HDM raises
entropy of the universe upto $\sim 10^{87}$.   During the phase of
accelerated expansion, temperature falls as $a(t)^{- 50/103}$.

It is found that one of the two types of  dynamical changes of the universe
are possible, beyond $3.7 t_0  = 50.69{\rm Gyrs}$, (i) decelerated expansion and (ii) contraction. If future universe expands with deceleration , it will expand for ever. But in the case of contraction, it will collapse by $247.73{\rm Gyrs}$.

As $dark$ $energy$ dominates over $dark$ $matter$, from the beginning upto the time $3.03 t_0$, no $cosmic$ $coincidence$ $problem$ arises in the proposed  scenario. 
Moreover, in the preceding section, some  other basic problems such as
creation of particles in the early universe, primordial nucleosynthesis and
structure formation in the late universe. In the proposed model, it is shown
that particles are created due to topological changes which is unlike the SMU,
where it is assumed that elementary particles are created at the time of
big-bang. It is also shown, in section 7.2, that process of primordial
nucleosynthesis goes very well in the proposed model predicting $\sim 23\%$
Helium-4 abundance by weight. In section 7.3, it is shown that ,in the late
universe, inhomogeneities in CDM grow exponentially with time.

Thus  the proposed model is able to provide possible solutions to many
cosmological problems with prediction for the future universe.

\bigskip

\centerline{\bf Appendix A}

\centerline{\underline{\bf Riccion and Graviton}}

\smallskip

From the action

$$ S = \int {d^4 x} {d^D y}  \sqrt{- g_{(4+D)}} \quad
\Big[ \frac{M^{(2+D)}R_{(4+D)}}{16 \pi
} + {\alpha_{(4+D)}} R_{(4+D)}^2 + $$
$$\gamma_{(4+D)} ( R_{(4+D)}^3   - \frac{6(D+3)}{D-2)}{\Box}_{(D+4)}R^2_{(D+4)})\Big], \eqno(A.1)$$
the gravitational equations are obtained as

$$ \frac{M^{(2+D)}}{16 \pi} (R_{MN} - {1 \over 2} g_{MN} R_{(4+D)} ) + {\alpha_{(4+D)}} H^{(1)}_{MN} + {\gamma_{(4+D)}} H^{(2)}_{MN} = 0,\eqno(A.2a)$$ 
where

$$ H^{(1)}_{MN} = 2 R_{; MN} - 2 g_{MN} {\Box}_{(4+D)} R_{(4+D)} - {1\over 2} g_{MN} R^2_{(4+D)} + 2 R_{(4+D)} R_{MN}, \eqno(A.2b)$$  
and
$$ H^{(2)}_{MN} = 3 R^2_{; MN} - 3 g_{MN} {\Box}_{(4+D)} R^2_{(4+D)} - \frac{6(D+3)}{(D-2)}\{- { 1 
\over 2} g_{MN}{\Box}_{(4+D)} R^2_{(4+D)}$$  $$+ 2{\Box}_{(4+D)}R_{(D+4)}R_{MN} + R^2_{; MN}\} - { 1 \over 2} g_{MN} R^3_{(4+D)} + 3 R^2_{(4+D)} R_{MN} . \eqno(A.2c) $$

Taking $g_{MN} = \eta_{MN} + h_{MN}$ with $\eta_{MN}$ being $(4+D)$-dimensional Minkowskian metric tensor components and $h_{MN}$ as small fluctuations, the equation for $graviton$ are obtained as

$${\Box}_{(4+D)} h_{MN} = 0   \eqno(A.3) $$
neglecting  higher-orders of $h$.

On compactification of $M^4 \otimes S^D$ to $M^4$, eq.(A.3) reduces to the equation for 4-dimensional $graviton$ as
$${\Box} h_{\mu\nu} + \frac{l(l+D-1)}{\rho^2}h_{\mu\nu} = 0   \eqno(A.4) $$
for the space time 

$$ d S^2 = g_{\mu\nu} d x^{\mu} d x^{\nu} - \rho^2 [ d\theta_1^2 + sin^2\theta_1 d\theta^2_2 + \cdots +
sin^2\theta_1 \cdots sin^2\theta_{(D-1)} d\theta^2_D.  \eqno(A.5)$$

The 4-dimensional graviton equation (A.4) is like usual 4-dimensional graviton equation ( the equation derived from 4-dimensional action) only for $l = 0$. Thus the massless graviton is obtained for $l = 0$ only.

As explained, in section 2, the trace of equations (A.2) leads to the $riccion$ equation

$$[{\Box} + {1 \over2} \xi R + m^2_{\tilde R} + \frac{\lambda}{3!} {\tilde R}^2]{\tilde R} + \vartheta  = 0,  \eqno(A.6a)$$
where

\begin{eqnarray*}
 \xi & = & \frac{D}{2(D+3)} +  \eta^2 \lambda R_D \\ m^2_{\tilde R} & = & - \frac{(D + 2) \lambda V_D}{16 \pi G_{(4+D)}} + \frac{D R_D}{2(D+3)} + \frac{1}{2} \eta^2 \lambda R_D^2 \\ \lambda  & = & \frac{1}{4(D+3)\alpha}, \\ \vartheta & = & - \eta \Big[- \frac{(D + 
2) \lambda M^{(2+D)}V_D}{16 \pi} + \frac{D  R_D^2}{4 (D+3)} + \frac{1}{6}
\eta^2 \lambda R_D^3 \Big].
\end{eqnarray*}

The graviton $h_{\mu\nu}$ has 5 degrees of freedom (2 spin-2 graviton, 2
spin-1 gravi-vector (gravi-photon) and 1 scalar). The scalar mode $f$
satisfies the equation

$$ {\Box} f + \frac{l(l + D - 1)}{\rho^2} f = 0 \eqno(A.7)$$
Comparison of eqs.(A.6) and (A.7) show many differences between scalar mode
$f$ of graviton and the riccion $(\tilde R)$ e.g. $\xi, \lambda$ and
$\vartheta$ , given by eqs.(A.6b,c,d,e), are vanishing for $f$, but
non-vanishing for $\tilde R$. Eq.(A.7) shows $(mass)^2$ for $f$ as 

$$ m_f^2 = \frac{l(l + D - 1)}{\rho^2} f , \eqno(A.8)$$
whereas $(mass)^2$ for $\tilde R$, given by eq.(A.6c), depends on $G_{4+D},
V_D $ and $R_D$ (given in section 2). $m_f^2 = 0$ for $l = 0,$ but $
m^2_{\tilde R}$ can vanish only when gravity is probed upto $\sim 10^{-33}
cm.$ As mentioned above, so far, gravity is probed only upto 1cm. Thus
$(mass)^2$ of riccion does not vanish. 

So, even though, $f$ and ${\tilde R}$ are scalars arising from gravity, both
are different. $Riccion$ can not arise without higher-derivative curvature
terms in the gravitational action, but $graviton$ can be obtained even from
Einstein-Hilbert action.

$${\Box} = \frac{1}{\sqrt{-g}} \frac{\partial}{\partial x^{\mu}}
\Big(\sqrt{-g} g^{\mu\nu} \frac{\partial}{\partial x^{\nu}} \Big) =
\eta^{\mu\nu}\frac{\partial}{\partial X^{\mu}\partial X^{\nu}  }, $$
where ${X^{\mu}}$ are locally inertial co-ordinates and $\eta^{\mu\nu}$ are
Minkowskian metric components. It shows that the scalar like operator ${\Box}$
has the same role on $\tilde R$ as it is for other scalar fields $\phi$ due to
principle of equivalence. 

\bigskip

\centerline{\bf Appendix B}

\centerline{\underline{\bf Bogoliubov transformations}}

\smallskip

In a Hilbert space, a sacalar $\Phi$ satisfying the Klein-Gordon equation can
be written as linear combinations

\begin{eqnarray*}
\Phi &=& {\sum_k} \Big[ A_k^{\rm in} \Phi_k^{\rm in} + A_k^{\dag{\rm in}}
\Phi_k^{*{\rm in}} \Big] \\ &=& {\sum_k} \Big[ A_k^{\rm out} \Phi_k^{\rm out}
+ A_k^{\dag{\rm out}} \Phi_k^{*{\rm out}} \Big],
\end{eqnarray*}
$$ \eqno(B.1a)$$
where $\Phi^{\rm out}_{k}$ and $\Phi^{\rm in}_{k}$ both belong to
the same Hilbert space. As a result, one obtains
$$ \Phi^{\rm out}_{k l m} = \alpha_{k} \Phi^{\rm in}_{k}  +
\beta_{k} \Phi^{*{\rm in}}_{k} , \eqno(B.1b)$$ 
where $\alpha_{k}$ and $\beta_{km}$ are Bogolubov coefficients
satisfying the condition 
$$|\alpha_{k}|^2  - |\beta_{k}|^2  = 1. \eqno(B.2)$$

The in- and out- vacuum states are defined as
$$A^{\rm in}_{k}|{\rm in}> = 0 = A^{\rm out}_{k}|{\rm out}>
.\eqno(B.3a,b)$$ 
Moreover,
$$A^{\rm out}_{k} = \alpha_{k} A^{\rm in}_{k}  +
\beta_{k} A^{*{\rm in}}_{k} , \eqno(B.3c)$$

The normalization condition for $\Phi$ is given as
$$ (\Phi_{k} , \Phi_{k }) = 1  = - (\Phi^*_{k 
} , \Phi^*_{k}),  (\Phi_{k} , \Phi^*_{k}) = 0 \eqno(B.4a,b,c)$$ 
where the scalar product 
is defined as
$$ (\Phi_{k } , \Phi_{k}) = -\int\sqrt{ - g_{\Sigma}}  {d{\Sigma}^{\mu}} [ \Phi_{k } (\partial_{\mu} \Phi^*_{k^{\prime}l^{\prime} m^{\prime}}) -
(\partial_{\mu} \Phi_{k}) \Phi^*_{k^{\prime}l^{\prime} m^{\prime}}] ,
\eqno(B.4d)$$ 
where $\Sigma$ is the 3-dim. hypersurface.

Connecting eqs.(B.1b) and (B.4a,b,c),
$$  \alpha_{k } = (\Phi^{\rm out}_{k } , \Phi^{\rm in}_{k }) \eqno(B.5a)$$
and
$$  \beta_{k } = - (\Phi^{\rm out}_{k } , \Phi^{*{\rm in}}_{k}) \eqno(B.5b)$$

\smallskip
\noindent\underline{(b) Spin-1/2 field}
\smallskip

The spin-1/2 field $\psi$, satisfies the Dirac equation
$$ \Big(i \gamma^{\mu} D_{\mu} - m_f ) \psi = 0,  \eqno(B.6a)$$
where $m_f$ is mass of $\psi$ and
$$D_{\mu} = \partial_{\mu} - \Gamma_{\mu} \eqno(B.6b)$$ 
with Dirac matrices in curved space-time
$$ \gamma^{\mu} = e^{\mu}_a {\tilde \gamma}^a,  \eqno(B.7a)$$
where $(a , \mu = 0,1,2,3)$ and $e^{\mu}_a$ are defined through 
$$e^{\mu}_a e^{\nu}_b g_{\mu\nu} = \eta_{ab}. \eqno(B.7b)$$
Here $\eta_{ab}$ are Minkowskian metric tensor components and $g_{\mu\nu}$
are metric tensor components in curved space-time. $c.c.$ stands for
complex conjugation.

Dirac matrices $\gamma^{\mu}$ in curved space-time satisfy the anti-commutation
rule \cite{sk96, ndb}
$$ \{ \gamma^{\mu} , \gamma^{\nu} \} = 2 g^{\mu\nu}  \eqno(B.7c)$$
and Dirac matrices ${\tilde \gamma}^a$ in Minkowskian space-time 
satisfy the anti-commutation rule
$$ \{ {\tilde \gamma}^a , {\tilde \gamma}^b \} = 2 \eta^{ab}.
\eqno(B.7d)$$ 
$\Gamma_{\mu}$ are defined as
$$\Gamma_{\mu} = - {1 \over 4} \Big(\partial_{\mu}e^{\rho}_a +
\Gamma^{\rho}_{\sigma\mu} e^{\sigma}_a \Big) g_{\nu\rho} e^{\nu}_b{\tilde
\gamma}^b {\tilde \gamma}^a.  \eqno(B.7e)$$ 

Further, $\psi$ can be decomposed as \cite{sk96}
\begin{eqnarray*}
 \psi &=& {\sum_{s=\pm1}}{\sum_k} \Big( b_{k,s}^{\rm in}
 \psi^{\rm in}_{I
(k,s)} + d^{\dagger {\rm in}}_{-k,-s} \psi^{\rm in}_{II (-k,-s)}
 \Big)\\
&=& {\sum_{s=\pm1}}{\sum_k} \Big( b_{k,s}^{\rm out}
 \psi^{\rm out}_{I 
(k,s)} + d^{\dagger{\rm out}}_{-k,-s}
 \psi^{\rm out}_{II (-k,-s)} \Big),
\end{eqnarray*}
$$ \eqno(B.8a,b)$$ 
as both in- and out-spinors belong to the same Hilbert space. The in- and
out-vacuum states are defined as 
$$  b_{k,s}^{\rm in}|{\rm in} > = d_{-k,s}^{\rm in}|{\rm in} > = 0
\eqno(B.9a,b)$$  
and
$$  b_{k,s}^{\rm out}|{\rm out} > = d_{-k,s}^{\rm out}|{\rm out} > = 0
\eqno(B.9c,d)$$

Bogoliubov transformations are given as \cite{sk96}
\begin{eqnarray*}
b_{k,s}^{\rm out} & = & b_{k,s}^{\rm in} \alpha_{k,s} +
d_{-k,-s}^{{\dagger}{\rm in}}\beta_{k,s} 
 \\ b_{k,s}^{{\dagger}{\rm out}} &
= &  \alpha^*_{k,s} b_{k,s}^{{\dagger}{\rm in}} +
 \beta^*_{k,s} d_{-k,-s}^{\rm 
in}  \\ d_{-k,-s}^{{\dagger}{\rm out}} & = &
 b_{k,s}^{\rm in} \alpha_{k,s} +
d_{-k,-s}^{{\dagger}{\rm in}}\beta_{k,s}  \\ 
 d_{-k,-s}^{\rm out} &
= &  \alpha^*_{k,s} b_{k,s}^{{\dagger}{\rm in}} +
 \beta^*_{k,s} d_{-k,-s}^{\rm 
in} .
\end{eqnarray*}
$$ \eqno(B.10a,b,c,d)$$
Connecting eqs.(B.8)-(B.10), one obtains \cite{skp}
$$|\alpha_{k}|^2 + |\beta_{k}|^2 = \sum_{s}|\alpha_{k.s}|^2 +
|\beta_{k.s}|^2 = 1 ,  \eqno(B.11)$$
where
$$\alpha_{k.s} = {\int_{\Sigma}} \sqrt{ - g_{\Sigma}} {d^3x} {\bar
\psi}_{I(k,s)}^{\rm out} {\tilde \gamma}^0 \psi_{I(k,s)}^{\rm in}
\eqno(B.12a)$$ 
and
$$\beta_{k.s} = {\int_{\Sigma}} \sqrt{ - g_{\Sigma}} {d^3x} {\bar
\psi}_{II(k,s)}^{\rm out} {\tilde \gamma}^0 \psi_{II(k,s)}^{\rm in}
\eqno(B.12b)$$ 

\bigskip

\centerline{\bf Acknowledgements}

\smallskip

Author is thankful to Inter University Centre for Astronomy and Astrophysics,
Pune for hospitality during my visit in January 2004, where idea for this
work came to my mind during 3 days seminar on brane-world. Thanks are also due
to  Varun Sahani for helpful suggestions. Useful discussions with Prof. K.P.Sinha, about riccions, is acknowledged gratefully.

\end{document}